\documentclass[11pt]{article}

\usepackage{graphicx}
\usepackage{amsfonts}
\usepackage{amsmath}
\usepackage{amsbsy}
\usepackage{subfigure}
\usepackage{natbib}
\usepackage{color}

\def\vv{{\bf {v}}}
\def\ww{{\bf {w}}}

\def\BB{{\bf {B}}}

\def\EE{{\bf {E}}}

\def\JJ{{\bf {J}}}

\def\PP{{\bf {P}}}

\begin{document}

\title{Theory of magnetic reconnection in solar and astrophysical plasmas\footnote{Published in {\it Phil.~Trans.~R.~Soc.~A}}}

\author{D.~I.~Pontin\footnote{e-mail: dpontin@maths.dundee.ac.uk}\\ Division of Mathematics, University of Dundee, Dundee, U.K. }

\date{}

\maketitle

\begin{abstract}
Magnetic reconnection is a fundamental process in a plasma that facilitates the release of energy stored in the magnetic field by permitting a change in the magnetic topology. In this article we present a review of the current state of understanding of magnetic reconnection. We discuss theoretical results regarding  the formation of current sheets in complex 3D magnetic fields, and describe the fundamental differences between reconnection in two and three dimensions. We go on to outline recent developments in modelling of reconnection with kinetic theory, as well as in the MHD framework where a number of new 3D reconnection regimes have been identified. We discuss evidence from observations and simulations of solar system plasmas that support this theory, and summarise some prominent locations in which this new reconnection theory is relevant in astrophysical plasmas.
\end{abstract}

\section{Introduction}
Magnetic reconnection is a fundamental process that is ubiquitous in astrophysical plasmas. The concept of reconnection is relevant in plasmas that are almost ideal -- i.e.~plasmas where the magnetic field is simply advected by the flow in the majority of the volume. Reconnection facilitates the release of energy stored in the magnetic field by permitting a breakdown in the magnetic connection between ideally-evolving plasma elements. As such, reconnection is universally accepted to be a key ingredient in  many astrophysical phenomena.

Historically, the concepts of reconnection are based on the original two-dimensional (2D), steady-state models which employed a magnetohydrodynamic (MHD) description of the plasma. As discussed below (\S\ref{2dsec}), these models failed in certain ways to describe the observed reconnection processes. Present research focusses in two main directions -- (i) investigation of two-fluid and particle effects not present in the original MHD description, and (ii) investigation of the reconnection process in three dimensions (3D), which turns out to be crucially different from the planar 2D case (\S\ref{fundsec}). 
In this article, we review recent advances in each of these avenues of research (\S\ref{2dsec},\ref{3dsec}).  
Our discussion of the theory of reconnection in 3D makes extensive use of the kinematic approximation, from which we may learn much about the underlying structure of the reconnection process. 
The purpose of this article is to highlight recent advances made in theory and modelling of reconnective processes on the Sun, and discuss their relevance to our understanding of astrophysical plasmas in general.
We therefore go on to describe  observations of energetic processes involving reconnection on the Sun in \S\ref{sunsec}, and discuss reconnection in astrophysical plasmas in general in  \S\ref{astrosec}.

\section{Fundamental properties of reconnection in 2D and 3D}\label{fundsec}
\subsection{Formation of current sheets}\label{jsheetform}
In order to understand the behaviour of many astrophysical plasmas, we must determine the locations where magnetic reconnection can facilitate the release of energy in the plasma. 
Whether in a collisional or collisionless plasma, reconnection requires the presence of a current sheet. 
We therefore require to understand (a) where current sheets form, and (b) how and where fast reconnection is triggered in those current sheets.
In some systems such as magnetospheres or pulsar winds, magnetic fields at different orientations are driven against one another on large scales, creating global-scale current sheets. In these cases, answering  question (b) is the key task. 
However, in, for example, the coronae of the Sun, stars, or accretions disks there are no global-scale current layers and the driving of the system is less direct, occurring at the magnetic field line footpoints on the star/disk surface. Here the question of how this photospheric driving maps eventually to the creation of coronal current sheets is not straightforward to answer, and issue (a) is most pertinent. Below we summarise some possible mechanisms and locations of current sheet formation in these indirectly-driven systems. We note that, in this review, we focus for the most part on the nature of individual reconnection events in laminar fields, and for the sake of brevity omit the effects of turbulence discussed by \cite{lazarian2012}.

\begin{figure}
\begin{center}
(a)\includegraphics[height=4.0cm]{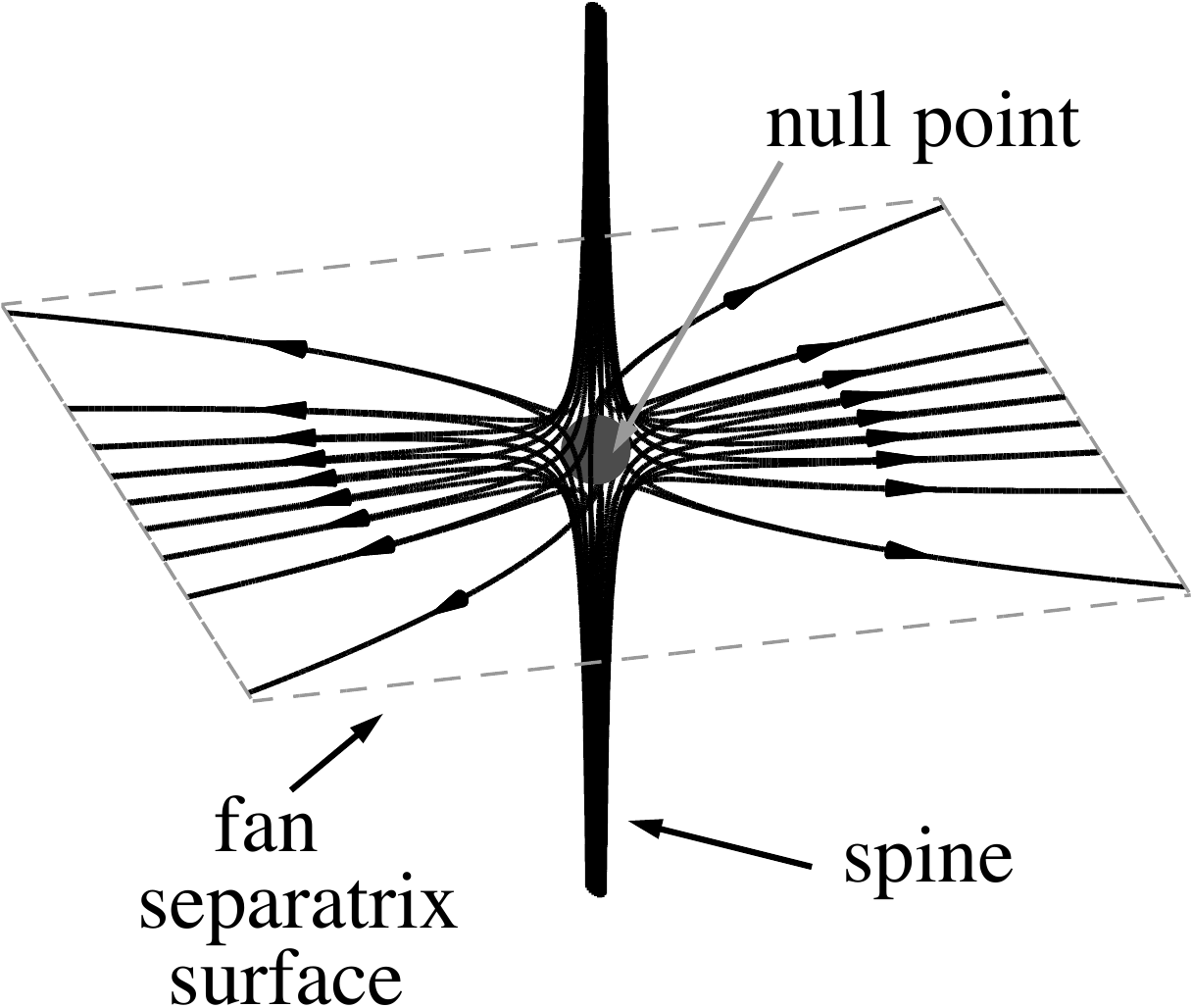}
(b)\includegraphics[height=5.0cm]{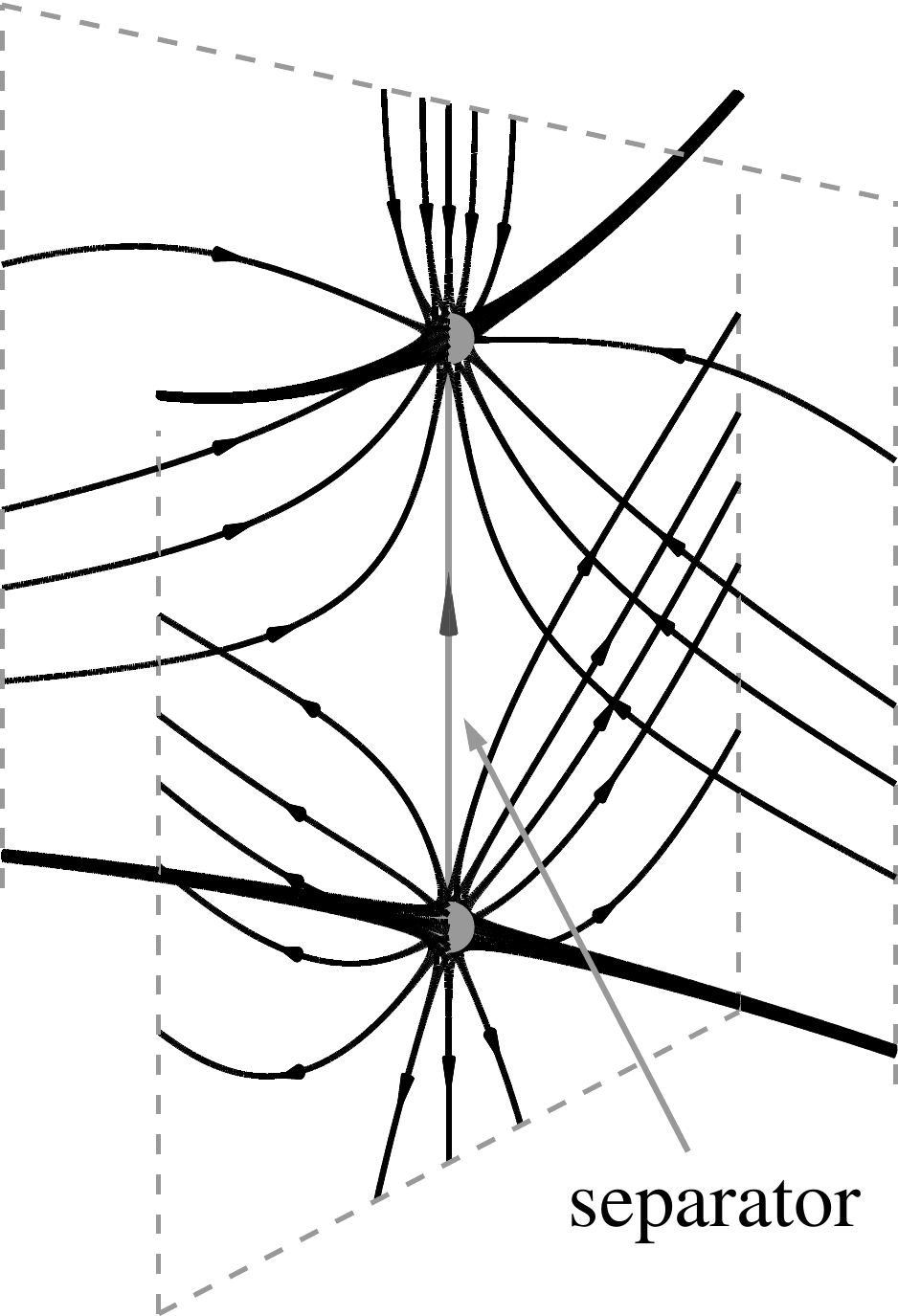}
\end{center}
\caption{Potential magnetic field line structure in the vicinity of (a) an isolated 3D null point, and (b) a generic fan-fan separator.}
\label{nullandsep}
\end{figure}

In 2D, it is well established that reconnection occurs at magnetic X-points, which are prone to collapse to form current layers \citep[e.g.][]{syrovatskii1971}. However, in 3D -- the case appropriate to the astrophysical environment -- the number of proposed sites of current sheet formation and reconnection is greatly increased. 
The structure of a magnetic field may be characterised in general by the mapping generated by the connectivity of magnetic field lines within the domain. A number of proposed sites of current sheet formation in 3D fields correspond to locations where this mapping is either discontinuous or has strong gradients. For example, the field line mapping is discontinuous at magnetic null points -- points in space at which $|\BB|=0$.
Since $\nabla\cdot {\bf B}=0$ these null points must be of hyperbolic type. Their structure \citep[categorised by][]{parnell1996} is characterised by a pair of field lines that asymptotically approach (or recede from) the null from opposite directions, forming the {\it spine} (or $\gamma$-) line, while field lines recede from (or approach) the null in a surface known as the {\it fan} (or $\Sigma$-) plane, see figure \ref{nullandsep}(a) (the spine and fan are called {\it invariant manifolds} in dynamical systems terminology). The fan surface is a {\it separatrix} surface -- it separates topologically distinct volumes of magnetic flux. Due to the discontinuous jump in field line connectivity, if one considers the implications of an ideal flow across the spine or fan in a kinematic model, singularities in the electric field result \citep{lau1990,priest1996}. It has therefore been proposed that in a dynamic regime, current sheets would form at these locations. The same arguments can be made in a field containing a pair of nulls joined by a separator line -- a field line that runs from one null point to the other  (as well as closed field lines -- more relevant in laboratory plasmas). 
 In dynamical systems theory the separator would be called a heteroclinic orbit -- it is defined by the transverse intersection of the fan planes of the two nulls (figure \ref{nullandsep}(b)) and is therefore  topologically stable. Indeed, the formation of current layers at 3D nulls and separators has been studied by, for example, \cite{bulanov1997,pontincraig2005,longcopecowley1996}.

In the absence of null points the field line mapping is continuous. However, it has been proposed that if sufficiently strong gradients are present in this mapping then intense current layers will form when the field is perturbed by plasma motions \citep[e.g.][]{longcope1994,priest1995}. These gradients in the connectivity are characterised by the so-called {\it squashing factor}, $Q$ \citep{titov2002,titov2007}, and regions with high values of $Q$ are usually termed {\it quasi-separatrix layers} (QSLs). The name stems from the fact that a true separatrix surface may be considered as the limit obtained when a QSL approaches zero thickness and  infinite $Q$. For a detailed review of these ideas see the paper by \cite{demoulin2006}.

At nulls, separators, and QSLs, current sheets may form dynamically in response to a driving of the system, or during a relaxation process. There are however other mechanisms of current sheet formation, including ideal MHD instabilities. \cite{browning2008} and \cite{hood2009} have followed the formation of current layers and subsequent energy release following the onset of the kink instability in a straight, line-tied flux tube in resistive MHD simulations. 
In addition, numerical modelling of the loss of stability of curved flux ropes, to either the kink \citep{kliem2004} or torus \citep{kliem2006} instability has implicated reconnection at a current sheet in a QSL beneath the flux rope as facilitating the subsequent eruption. Furthermore,  \cite{boozer2005}  has proposed that neighbouring magnetic field lines generically separate exponentially, and that this exponentially increasing separation may also lead to exponentially large currents and thus reconnection.

A question that naturally arises is: do the structures discussed above (nulls, separators, QSLs) exist in astrophysical plasmas? New observations and theoretical studies suggest that they are abundant. Analysis of {\it in-situ} observations by the Cluster mission shows the presence of multiple 3D null points located in the current sheet of the Earth's magnetotail \cite[e.g.][]{xiao2007}, while clusters of nulls (expected to be joined by separators) have been found in the global magnetosphere simulations of \cite{dorelli2007}.
The absence of magnetic field measurements in the solar corona renders direct detection of these features impossible there at present. However, increasingly detailed magnetograms at the level of the photosphere permit  extrapolation of the field into the corona. Such extrapolations in quiet sun regions contain an abundance of 3D nulls, with a high density at chromospheric levels that falls off exponentially with height \citep[e.g.][]{regnier2008,longcope2009}. Furthermore, both nulls and QSLs have been inferred to be present in many flaring and erupting active regions, as discussed later.

\subsection{Evolution of magnetic flux during reconnection}

Under what conditions does magnetic reconnection occur in 3D? To answer this question we first require a definition of reconnection in 3D. Here we follow the most general framework of {\it general magnetic reconnection} \citep{schindler1988,hesse1988}, within which reconnection is defined by a breakdown of magnetic field line and flux conservation, or in other words a breakdown in the magnetic connection between ideally-evolving plasma elements. This has been shown to occur in 3D in general when a component of the electric field parallel to the magnetic field ($E_\|$) is spatially localised in all three dimensions. The change of connectivity, or reconnection rate, is quantified by the maximal value (over all field lines) of
\begin{equation}\label{recratedef}
\Phi=\int E_\| ds
\end{equation}
where the integral is performed along magnetic field lines from one side of the non-ideal region (region within which $E_\| \neq 0$) to the other.

In order to discuss further the properties of 3D magnetic reconnection, we first introduce the concept of {\it topological equivalence}. We call two magnetic fields on the same domain topologically equivalent if one can be obtained from the other via some (smooth) ideal evolution, or in other words one can be transformed into the other by means of a smooth (continuously differentiable) deformation. If the field lines cross the boundaries of the domain we require the velocity on the boundaries to vanish\footnote{The concept can be generalised to allow for topological equivalence based on a prescribed non-zero boundary motion, but we omit this here for simplicity.}. This implies that topologically equivalent fields preserve all connections of field lines from one boundary to another and also all linkages or knottedness of field lines within the volume. As a consequence  the number and types of  magnetic null points are, for example, also preserved. All magnetic fields which are topologically equivalent in this sense form an equivalence class and we say that they have {\it the same topology}. 
Thus the above definition of general magnetic reconnection can be interpreted as follows. If, during some evolution of the plasma, the magnetic connection between plasma elements is not preserved, then by definition the magnetic field changes its topology -- and if this change is due to a local non-ideal evolution (as opposed to e.g.~a global diffusion) we call this magnetic reconnection. For a detailed discussion of the relationship between magnetic topology and magnetic reconnection, the reader is referred to the papers by \cite{hornig1996,hornig2007a,yamada2010}.

It is now clear that the fundamental properties of 3D reconnection are crucially different from the simplified 2D picture. These can be understood by considering the equation
\begin{equation}\label{idealev}
\frac{\partial {\bf B}}{\partial t}- \nabla\times (\ww \times \BB)={\bf 0},
\end{equation}
which describes the ideal evolution of a magnetic field $\BB$, where ${\bf w}$ is a flux-conserving velocity or flux transport velocity (which in ideal MHD is simply the fluid velocity ${\bf v}$). If for a given magnetic field evolution a smooth flow ${\bf w}$ exists then the magnetic flux is {\it frozen into} the flow $\ww$, and the topology of the magnetic field is preserved -- this being guaranteed by the condition that ${\bf w}$ be smooth and continuous. (Strictly speaking, the topology is still preserved if there is a non-zero term of the form $\lambda\BB$ on the right-hand side of equation (\ref{idealev}), where $\lambda$ is some scalar field.) In 2D $\EE\cdot\BB$ is identically zero everywhere, and the conditions on the solution of equation (\ref{idealev}) are straightforward. A flux transporting flow exists everywhere, but is singular at null points of the magnetic field if ${\bf E}\neq{\bf 0}$ there.  However, in 3D reconnection by definition $\EE\cdot\BB\neq 0$ (see above), and the conditions under which magnetic topology conservation, field line conservation, and magnetic flux conservation hold are much more subtle -- the reader is referred to the papers by \cite{schindler1988,hornig1996,hornig2001,hornig2007a}.

The singularity of $\ww$ at 2D magnetic X-points is a signature of the fact that the reconnection process involves magnetic field lines being cut and rejoined at the X-point. In other words, the connectivity of each field line changes in a discontinuous manner when it passes through the separatrices and null. Since $\ww$ is smooth and continuous everywhere except at the X-point, field lines evolve as if they are reconnected at this point {\it only}. So the reconnection of magnetic field lines occurs in a one-to-one pairwise fashion at a single point. 

\begin{figure}
\begin{center}
\includegraphics[width=4.1cm]{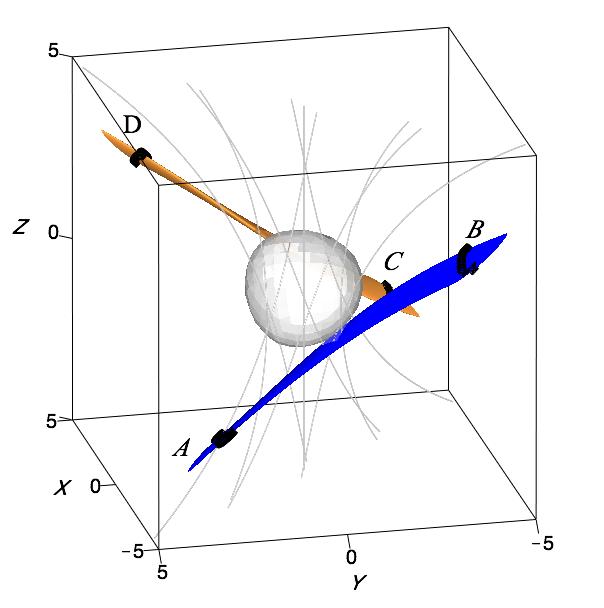}
\includegraphics[width=4.1cm]{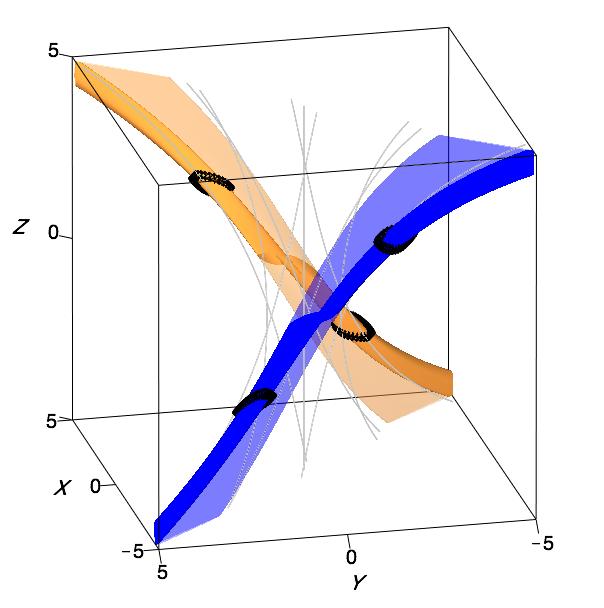}
\includegraphics[width=4.1cm]{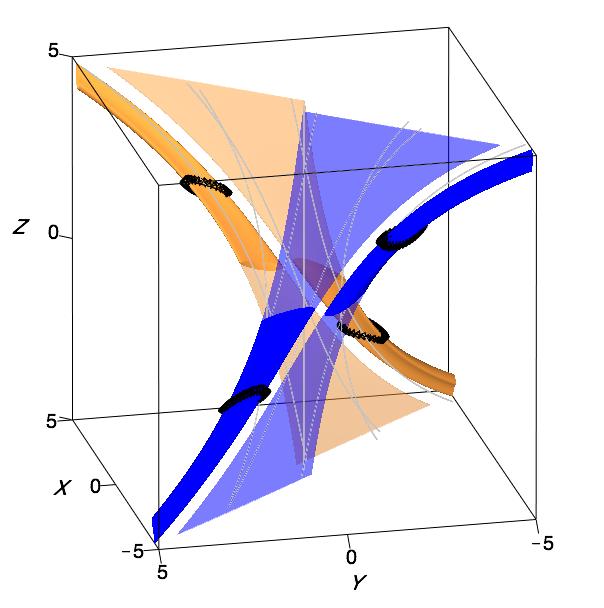}
\includegraphics[width=4.1cm]{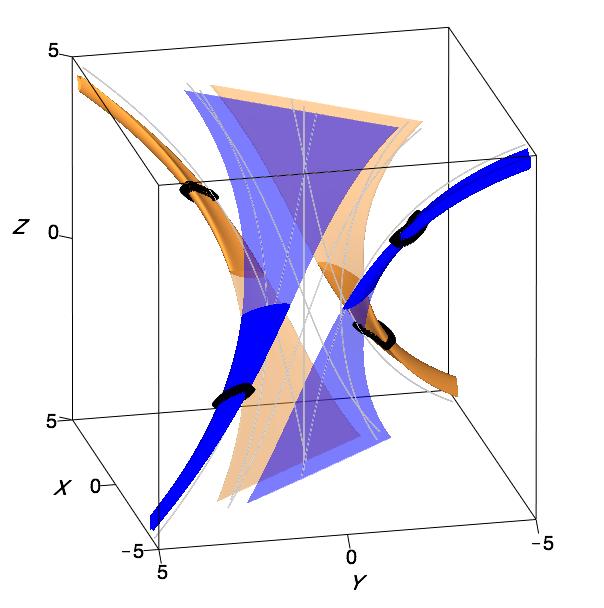}
\includegraphics[width=4.1cm]{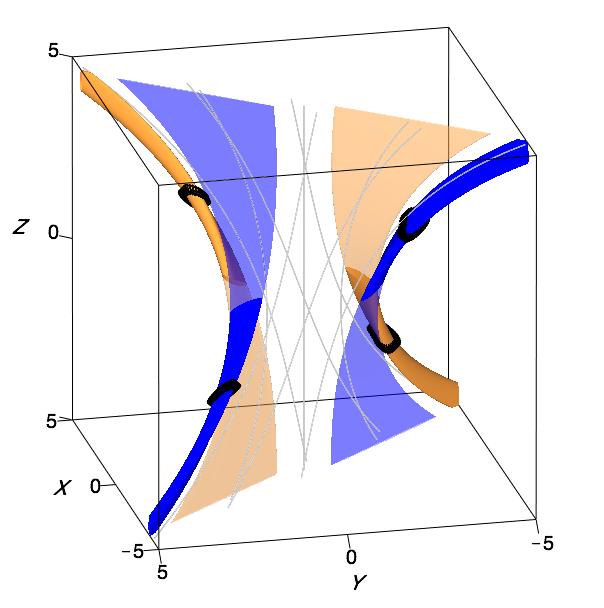}
\includegraphics[width=4.1cm]{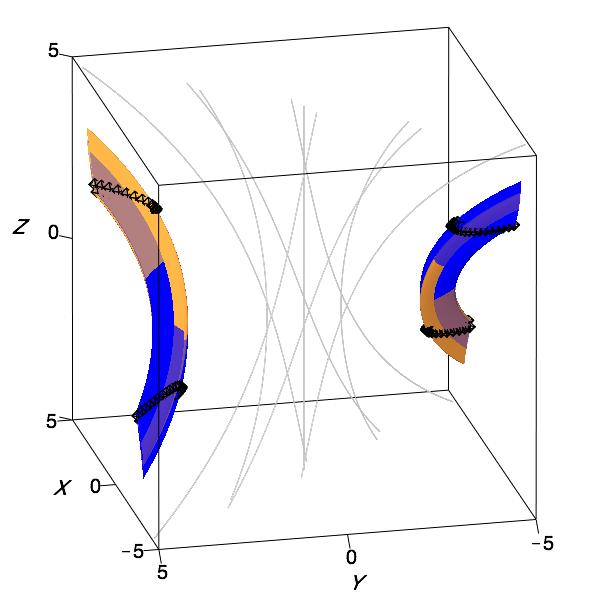}
\end{center}
\caption{Reconnection of representative flux tubes in the magnetic field $\BB=(y, k^2 x,1)$, with $k=1.2$, traced from ideal comoving footpoints (marked black). The solid sections move at the local plasma velocity (outside the non-ideal region), while the transparent sections correspond to field lines that pass through the non-ideal region.  The shaded surface in the first frame shows the localised non-ideal region. (After \cite{pontin2011b}.)}
\label{tubeflip}
\end{figure}

Perhaps surprisingly, it turns out that none of the above properties of 2D reconnection carry over into 3D. In the presence of a localised non-ideal region in 3D, the evolution of the magnetic flux has the following properties:
\begin{itemize}
\item
A flux transporting velocity $\ww$ {\it does not} exist for the flux threading the diffusion region (for a proof, see \cite{priesthornig2003}). 
\item
As a result, if one follows magnetic field lines from footpoints comoving in the ideal flow, they appear to split {\it as soon as they enter the non-ideal region}, and their connectivity changes {\it continually and continuously} as they pass through the non-ideal region (see figure \ref{tubeflip}). In other words, between any two neighbouring times $t$ and $t+\delta t$, {\it every} field line threading the non-ideal region experiences a change in connectivity. 
\item
Magnetic field lines are {\it not} reconnected in a one-to-one fashion. Consider two field lines that are about to enter the diffusion region, one of which connects plasma elements labelled A and B,  while the other connects plasma elements C and D (as in figure \ref{tubeflip}). Then if the field lines are chosen such that after reconnection A is connected to C, then B will {\it not} be connected to D. 
\end{itemize}

The above properties are demonstrated in figure \ref{tubeflip}. The plots are based on the steady-state kinematic solution of \cite{hornig2003}, with $\BB=(y, k^2 x,1)$ and an anomalous resisitivity localised around the origin.  Note that in this model the resistivity has been localised in order to make the equations tractable -- however, the above described topological properties of the flux evolution are not dependent on this localisation, and are still present when the non-ideal region is self-consistently localised through the formation of a localised current layer (as discussed later). Representative field lines are traced from four cross sections comoving in the ideal flow, chosen such that at the initial time they form a pair a flux tubes. The flux tubes immediately begin to split when they enter the non-ideal region, with field lines from cross-sections A and B (say) no longer being coincident. In frames 2-5 of the figure, the apparent `flipping'  of field lines is demonstrated. The solid sections of the flux tubes are traced from ideal comoving footpoints (marked black) and move at the local plasma velocity (outside the non-ideal region), while the transparent sections correspond to field lines traced into and beyond the non-ideal region, and appear to flip past one another at a velocity that is different from the local plasma velocity (until they exit the non-ideal region). Note that while in the initial state we began with two flux tubes, after reconnection the four cross-sections do not match up to form two unique flux tubes (in contrast to the 2D case) -- see the bottom-right frame of the figure. 

The above 3D reconnection properties hold regardless of the structure of the magnetic field in the vicinity of the reconnection region (for example the presence of nulls and separators). For further details, the reader is referred to the papers by \cite{schindler1988,hornig2001,hornig2007a,priesthornig2003}. We will discuss reconnection in 3D further in \S\ref{3dsec}.

\section{Reconnection in two dimensions}\label{2dsec}
\subsection{The MHD limit}
The first quantitative model for a reconnection process in a current layer was put forward by \cite{sweet1958} and \cite{parker1957}. In the model, anti-parallel magnetic fields are brought together at a current sheet of length $2L$ and thickness $2l$, with the magnetic energy being converted into kinetic energy associated with the outflow jets and heat via Joule dissipation. The intention of Sweet and Parker was to explain energy release in solar flares. However, analysis showed that the rate at which flux could be reconnected (and energy subsequently released) was many orders of magnitude too slow to account for a typical solar flare. The principal reason for this is that, owing to mass conservation, the reconnection rate scales with the ratio of the sheet dimensions $l/L$, which itself scales as $\sqrt{\eta}$. This poses a major problem in astrophysical plasmas where the plasma resistivity $\eta$ is thought to be very small. 

This problem led \cite{petschek1964} to propose an alternative mechanism. In this model the current layer no longer has a macroscopic length $L$, but rather standing slow-mode shock waves emanate from the ends of the current layer, and much of the energy conversion occurs within these shocks. Different reconnection rates are possible depending on the exact configuration assumed, and the maximum allowable reconnection rate scales as $\ln({\eta})$. Due to the weak scaling with $\eta$, and the consequently much higher (than Sweet-Parker) reconnection rate for astrophysical plasma parameters, Petschek's model is described as a `fast' reconnection model. 

Owing to the ground-breaking nature of the two models discussed above, to this day new results in 2D reconnection modelling are still interpreted in their terms. In particular, models or simulations which lead to current layers with macroscopic or system-scale length are described as `Sweet-Parker-like', while any system with a (quasi-)steady current layer which remains of microscopic length is usually termed `Petschek-like'. Note that many other modifications to these early models were subsequently proposed, see e.g.~\cite{priest2000,biskamp2000} for reviews.

\subsection{Two-fluid and kinetic effects}
For some time it appeared that the problem of fast reconnection was solved. However, it has since been shown via MHD simulations that in fact Petschek-like solutions are not observed in an MHD evolution \citep[e.g.][]{biskamp1986}, but rather, Sweet-Parker-like current layers are found -- at least in simulations with spatially-uniform resistivity. Since this realisation, a great deal of effort has been spent on understanding how Petschek-like fast reconnection can work. This has largely involved the introduction of extra physics not included in the MHD description.
However, it is worth noting that recent work by \cite{baty2009a,baty2009b} shows that Petschek-like fast reconnection configurations can be achieved when only very weak fluctuations are present in either the plasma resistivity or viscosity.

In a particle plasma description, the generalised form of Ohm's law that is appropriate can be written 
\begin{equation}\label{generalohm}
\EE+\vv\times\BB=\eta \JJ  - \frac{\JJ\times\BB}{ne}-\frac{1}{ne}\nabla\cdot\underline{\PP}_e + \frac{m_e}{ne^2}\left( \frac{{\partial \JJ}}{\partial t}+\nabla \cdot ({\bf v}\JJ+\JJ\vv) \right)
\end{equation}
where $\JJ$ is the current density, $n$ is the particle number density, $e$ the electron charge, $m_e$ the electron mass, and $\underline{\PP}_e$ the electron stress tensor. The first term on the right-hand side is now interpreted as describing electron-ion collisions,  the second term  is the Hall electric field, while the final term describes the role of electron inertia. 

Now, for fast reconnection in 2D, one requires a suitably large electric field $\EE$ in the current sheet. It is clear that the Hall electric field cannot balance the reconnection electric field at the null -- rather the electron pressure or inertia terms must be responsible. However, results from a set of simulations \citep{birn2001} seemed to demonstrate that in any system which includes Hall physics -- and associated whistler waves -- fast reconnection occurs. Thus it has been proposed that the essential physics required to set up a Petschek-like current sheet is the Hall effect. Indeed,  \cite{drake2008} have argued that the whistler waves are essential for limiting the length of the diffusion region. \cite{cassak2005} proposed that a `catastrophic' transition between Sweet-Parker-like and Hall reconnection regimes can account for fast triggering of the reconnection process.

However, this issue is still controversial, one reason being that recent simulations  have demonstrated fast reconnection in electron-positron plasmas, in which by definition the Hall term is zero. In these studies, off-diagonal terms in the electron pressure tensor have most often been proposed as providing the mechanism responsible for the dissipation \citep[e.g.][]{bhesso2005,daughton2007}. Furthermore, one critical unknown property of kinetic simulation results is how they will scale to large systems, i.e.~how they can match onto the external MHD region. \cite{daughton2007} and \cite{daughton2009} have argued that the fast reconnection solutions obtained in typical Hall-dominated reconnection simulations remained stable due to symmetric boundary conditions in the outflow direction. They claim that when very large systems (or systems with open boundary conditions) are considered the current layer will always grow in length towards a Sweet-Parker-like configuration. Fast reconnection is then obtained due to the loss of stability of the current sheet which continually spawns plasmoids or `secondary islands', this process ensuring that the current layer never becomes so large that the reconnection process is excessively slowed down \citep[e.g.][]{loureiro2009,bhattacharjee2009,uzdensky2010}.

We have given here only the briefest of introductions to current research into the structure of the current layer in 2D reconnection. Excellent recent reviews can be found in \cite{birn2007,zweibel2009,yamada2010}.

\section{Three-dimensional reconnection regimes}\label{3dsec}
In three dimensions there are a number of structures in the magnetic field that may harbour current sheets, as discussed in \S\ref{jsheetform}. The different structures of the magnetic field lead to different `regimes' of 3D reconnection. These are summarised below -- a more detailed review is available in the paper by \cite{pontin2011b}.

\subsection{Reconnection without null points}\label{nonnullrec}
As discussed above, one location that magnetic reconnection may occur in 3D is in current layers that are not associated with magnetic nulls. The continuous change of connectivity of field lines traced from comoving footpoints has led to such reconnection being termed {\it magnetic flipping} \citep{priest1992} or {\it slip-running reconnection} (if the virtual flipping velocity exceeds some threshold, \cite{aulanier2006}). 

A major step in understanding the properties of 3D non-null reconnection was made by \cite{hornig2003}. They considered the kinematic resistive problem (i.e.~neglecting the equation of motion),  imposing a steady-state magnetic field and plasma resistivity and solving Ohm's law for the electric field and plasma velocity perpendicular to $\BB$ ($\vv_\perp$) via
\begin{equation}\label{kinsol}
\Phi=\int \eta \JJ\cdot\BB \, ds, \qquad \EE=-\nabla\Phi, \qquad \vv_{\perp} = \frac{(\EE-\eta\JJ)\times\BB}{B^2},
\end{equation}
(the parallel component of $\vv$ being arbitrary).
The magnetic field was taken to be $\BB={B_0} (y,k^2 x,1)/L$. Since this field is linear the current is uniform, and since the authors' aim was to study an isolated 3D reconnection process -- the generic case in astrophysical plasmas -- the imposed resistivity was localised around the origin, in order to fully localise the non-ideal region (defined as the non-zero product of $\eta {\bf J}$).  

Solving equations (\ref{kinsol}) shows that the plasma flow required to maintain this steady state configuration is a counter-rotational flow. More precisely, the flow is confined to field lines which thread the diffusion region, $D$, with plasma above and below $D$ (with respect to the direction of $\BB$) rotating in opposite senses. One can demonstrate that this is a necessary property of the solution that follows directly from the presence of a 3D-localised parallel electric field within a region of non-vanishing magnetic field \citep{hornig2003}. That is, it is independent of the particular choice of spatial profiles of $\BB$ and $\eta$, or indeed the fact that $\eta$ rather than $\JJ$ is localised (a localisation of the current would be a more physically plausible way to localise $E_\|$, but is not compatible with the method of solution). 

As a consequence of the counter-rotational flows, field lines followed from the ideal region above and below $D$ undergo a `rotational slippage' with respect to one another, which is quantified by the reconnection rate calculated via equation (\ref{recratedef}). It is worth emphasising that this characteristic flow structure for 3D non-null reconnection is very different to the classical 2D reconnection picture in which the characteristic flow structure is of stagnation type. The counter-rotational flows are a signature of the helicity production (decay) in 3D reconnection: if one writes down an evolution equation for the magnetic helicity then $\EE\cdot\BB$  appears as a source term. 
The solution of \cite{hornig2003} allows for the addition of an ideal flow via the constant function of integration in the integral in equation (\ref{kinsol}). 
The authors considered the effect of adding a flow with a hyperbolic structure in the $xy$-plane to transport magnetic flux into and out of the diffusion region. The result is that field lines are brought into the non-ideal region, are split apart by the counter-rotational flows, and exit differently connected in opposite quadrants of the flow. The evolution of a particular pair of flux tubes for one of these solutions is shown in figure \ref{tubeflip}.

The properties of the above solution, including the presence of counter-rotational flows, were verified in a resistive MHD simulation \citep{pontingalsgaard2005}. A further instance in which reconnection in the absence of null points is seen in simulations is in 3D studies of Parker's `topological dissipation' hypothesis \citep[e.g.][]{galsgaard1996,rappazzo2008}. While the topology of the magnetic field in the vicinity of the reconnection sites is not determined in these studies, the presence of a strong background field ensures the absence of null points. It is worth noting that the local structure of the magnetic field at a non-null reconnection processes can be either hyperbolic or elliptic \citep[e.g.][]{wilmotsmith2010}. Another notable instance of non-null 3D reconnection simulations is the study of \cite{linton2001} who investigated the interaction of twisted magnetic flux tubes in an otherwise field-free environment, and discovered different possible interactions (`merge', `bounce', `tunnel' and `slingshot') depending on the relative orientations of the tubes.

\subsection{Reconnection at null points}
\subsubsection{Kinematic models}
Early models for 3D null point reconnection were proposed by \cite{priest1996} who considered  the ideal kinematic limit and a current-free magnetic null. However, \cite{pontin2004,pontinhornig2005} showed that the possible magnetic flux evolutions are very different when a localised non-ideal region is included around the null point. They performed a similar kinematic analysis to that of \cite{hornig2003} as described above, and found that the nature of the magnetic reconnection is crucially dependent on the orientation of the electric current at the null. 

\begin{figure}
\begin{center}
\includegraphics[width=4.0cm]{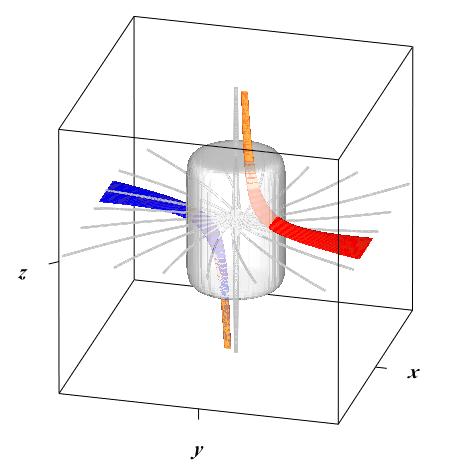}
\includegraphics[width=4.0cm]{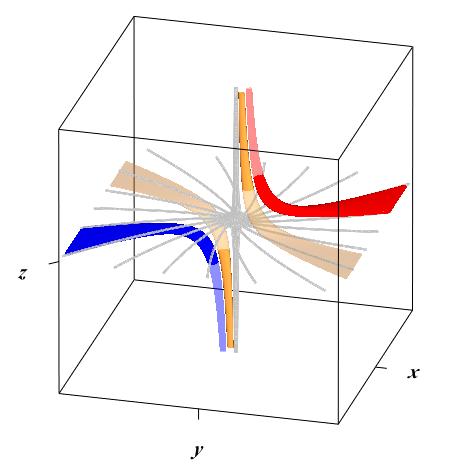}
\includegraphics[width=4.0cm]{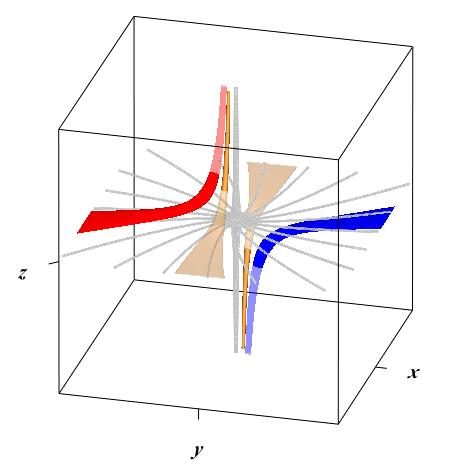}
\end{center}
\caption{Reconnection of representative flux tubes in the null point magnetic field $\BB=(r, r/2,-2z)$ in cylindrical polar coordinates, corresponding to current directed parallel to the spine. A localised diffusion region (shaded surface, top-left) is present around the null point. Flux tubes are traced from four ideal comoving footpoints, with their extensions that pass through the diffusion region rendered as transparent, as in Figure \ref{tubeflip}.}
\label{tube_spinealigned}
\end{figure}

If the current is directed parallel to the spine of the null, then there are counter-rotational flows, centred on the spine \citep{pontin2004}. The change of connectivity that results from the reconnection process therefore takes the form of a rotational slippage similar to that  described above (see figure \ref{tube_spinealigned}). Importantly, there is no flux transport across either the spine or fan. The reconnection rate quatifies the difference between the rate of (rotational) flux transport in the ideal region on either side of the diffusion region.

\begin{figure}
\begin{center}
\includegraphics[width=4.2cm]{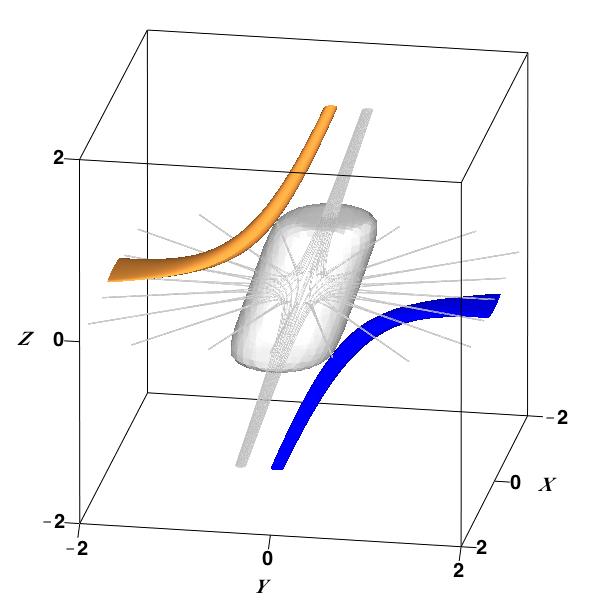}
\includegraphics[width=4.2cm]{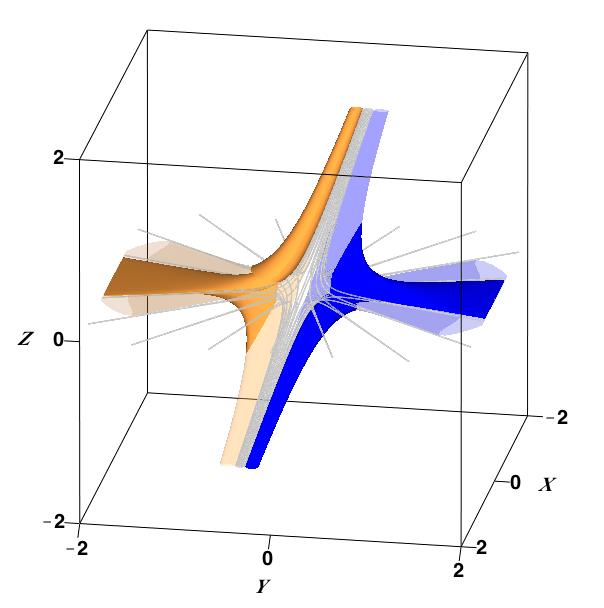}
\includegraphics[width=4.2cm]{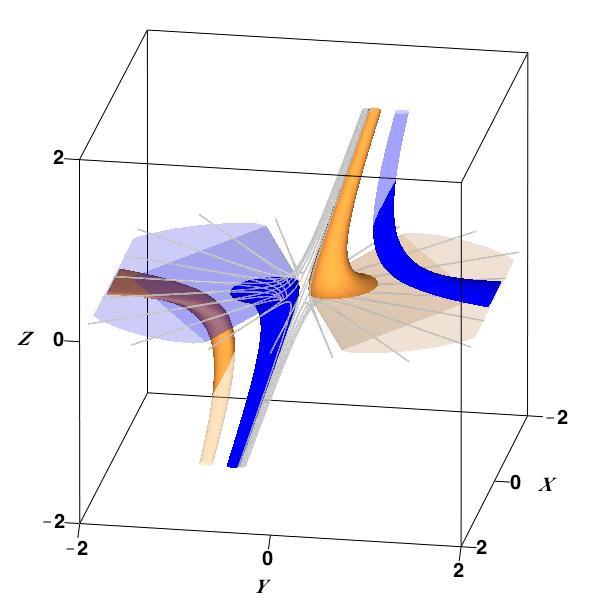}
\includegraphics[width=4.2cm]{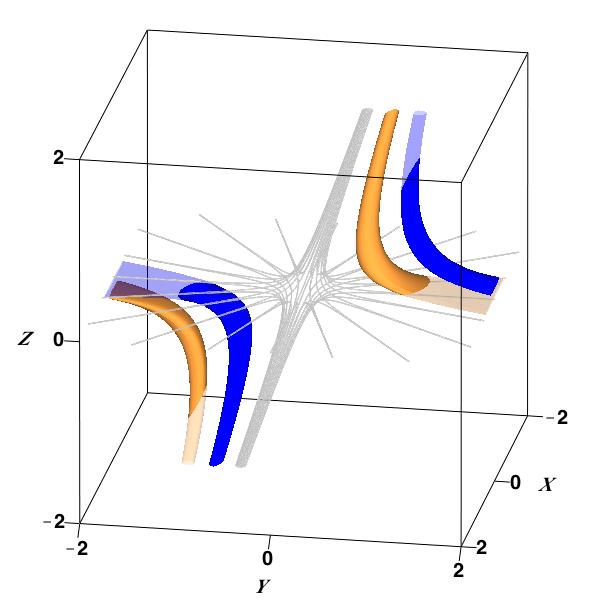}
\end{center}
\caption{Reconnection of representative flux tubes in the null point magnetic field $\BB=(x,y-z,-2z)$, corresponding to current directed parallel to the fan plane. A localised diffusion region (shaded surface, top-left) is present around the null. Flux tubes are traced from four ideal comoving footpoints, with their extensions that pass through the diffusion region rendered as transparent, as in Figure \ref{tubeflip}. (After \cite{pontin2011b}.)}
\label{tube_fanaligned}
\end{figure}

By contrast, when the current at the null is directed parallel to the fan surface, plasma flows across both the spine and fan of the null, transporting flux both through/around the spine line, and across the fan separatrix surface \citep{pontinhornig2005}, as shown in figure \ref{tube_fanaligned}. In this case, the reconnection rate can be shown to quantify the rate at which magnetic flux is transported across the separatrix surface in the ideal region---an interpretation that more closely resembles the 2D picture.

These kinematic solutions suggest two main modes of magnetic reconnection at 3D nulls. However, in these models the non-ideal region was artificially localised. The question still remains as to what types of current concentrations form self-consistently at 3D nulls in the dynamic regime. This has been investigated in a series of numerical simulations \citep{rickard1996,galsgaardpriest2003,pontingalsgaard2007,pontinbhat2007a}. The results have led \cite{priest2009} to propose a new categorisation of 3D null point reconnection regimes, as follows.

\subsubsection{Torsional spine and fan reconnection}
\cite{rickard1996} and \cite{pontingalsgaard2007} investigated the propagation of disturbances towards symmetric 3D null points (where the fan eigenvalues are equal). In both studies, a general disturbance was decomposed into rotations (in planes perpendicular to the spine) and shearing motions. Rotational motions are found to propagate along field lines, and accumulate around the spine line or fan plane. The locations of the spine and fan themselves remain undisturbed from their orthogonal potential configuration. These results are analogous to the properties of Alfv{\' e}n wave propagation towards 2D X-points, summarised by \cite{mclaughlin2011}. Due to the hyperbolic geometry of the magnetic field, the current intensifies as the length scales perpendicular to the spine or fan become shorter. The intensification ceases once these length scales become sufficiently short that diffusion becomes important.

\begin{figure}
\begin{center}
(a)\includegraphics[width=5cm]{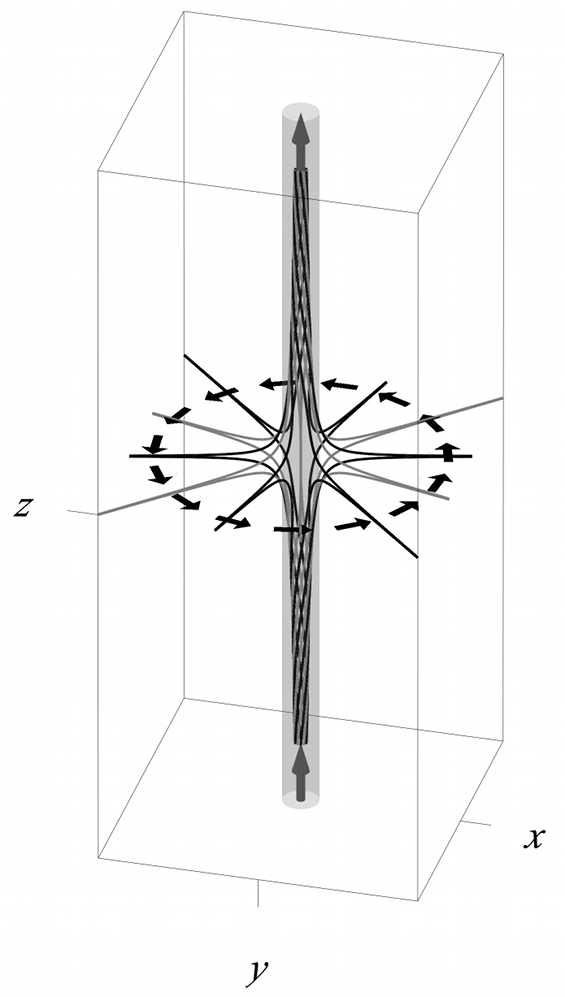}
(b)\includegraphics[width=6cm]{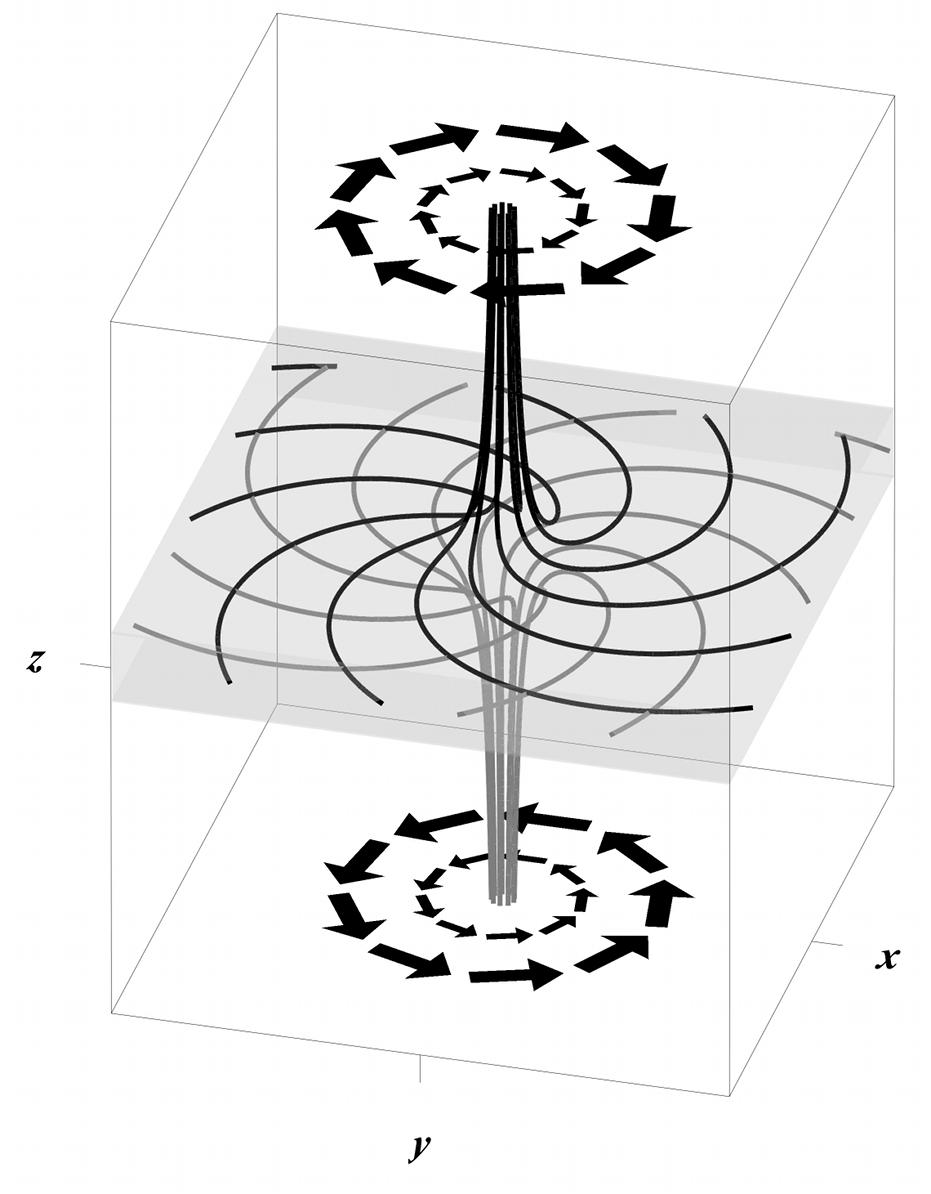}
\end{center}
\caption{Schematic diagrams of (a) torsional spine and (b) torsional fan reconnection. Black and grey lines are magnetic field lines, the shaded surfaces are current density isosurfaces, grey arrows indicate the direction of the current flow, while black arrows indicate the driving plasma  velocity. (After \cite{pontin2011b}.)}
\label{torrec}
\end{figure}

{\it Torsional spine reconnection} occurs in response to a rotational disturbance of field lines in the fan. The disturbance propagates to the spine, around which an extended tube of current forms. This current tube is generated by a twisting of the magnetic field lines locally around the spine line, and as such the current vector is directed parallel to the axis of the tube, i.e.~parallel to the spine, see figure \ref{torrec}(a). Due to the orientation of the current, the magnetic reconnection that occurs within the current layer takes the form of a rotational slippage, as discussed above. 

{\it Torsional fan reconnection} occurs in response to a rotational disturbance of field lines around the spine. The perturbation propagates as a helical Alfv{\' e}n wave towards the fan, where a planar current layer develops \citep{galsgaardpriest2003}, see figure \ref{torrec}(b). While away from the null the current is dominated by its components parallel to this plane, it flows through the null parallel to the spine. Therefore the reconnection again involves a rotational slippage of magnetic flux. 

\subsubsection{Spine-fan reconnection}
\begin{figure}
\begin{center}
\includegraphics[width=6cm]{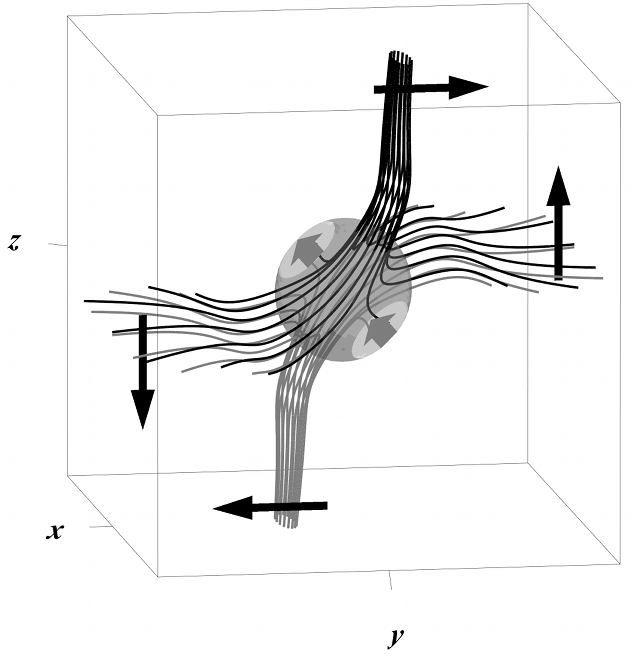}
\end{center}
\caption{Schematic diagram of spine-fan reconnection at an isolated null. Black and grey lines are magnetic field lines, the shaded surface is a current density isosurface, the grey arrows indicate the direction of the current flow, while the black arrows indicate the driving plasma  velocity. (After \cite{pontin2011b}.)}
\label{spinefan}
\end{figure}

The torsional spine and torsional fan reconnection modes discussed above require a rather organised rotational driving motion -- and it is thus anticipated that the most common regime of reconnection to occur at 3D nulls is the {\it spine-fan reconnection} mode. It occurs within a current sheet that is localised in all three dimensions around the null, such current concentrations being found to form when a shear disturbance of {\it either} the spine {\it or} the fan occurs \citep{pontinbhat2007a,galsgaard2011a}. 
The current layer at the null is formed by a local collapse of the magnetic field -- the spine and fan collapse towards one another, with the current sheet locally spanning them both, as depicted in figure \ref{spinefan}. Projected onto the plane in which the collapse takes place the spine, fan, and current layer together form a Y-type structure. The current flows through the null perpendicular to this plane, and thus parallel to the fan surface. As the null point collapses, magnetic flux is transported through {\it both} the spine {\it and} the fan, as predicted by the kinematic model due to the current orientation. \cite{alhachami2010} have demonstrated that the dimensions of the diffusion region and reconnection rate are strongly dependent on the degree of symmetry of the initial null point field.

It is worth noting the relation of the spine-fan reconnection regime to the steady-state mathematical models proposed by \cite{craigetal1995} and \cite{craig1996}. These `reconnective annihilation' solutions are constructed by super-imposing 1D or 2D disturbances consisting of infinite, straight field onto a background potential null, and therefore involve current layers that extend to infinity along either the spine or the fan. The solutions involving a planar current layer in the fan have been demonstrated to be dynamically accessible \citep{craig1998}. However, in order to maintain the planar nature of the current layer, a large pressure gradient is required within the layer. It has been shown in MHD simulations that when the incompressibility condition is relaxed the pressure gradient is not able to balance the Lorentz force acting within the current layer, whereupon the magnetic field collapses to form a  current sheet localised at the null as described above \citep{pontinbhat2007b}. 

\subsection{Reconnection at separator lines}\label{sepsec}
The form of current layers at separator  lines has been investigated by \cite{longcopecowley1996}, while \cite{longcope1996} has proposed that currents will naturally focus at separators during relaxation processes in the solar corona. Current accumulation at separators has been demonstrated in various numerical simulations \citep{galsgaard1997,haynes2007}. Early kinematic models in current-free magnetic fields predicted  a simple cut-and-paste of field line pairs at the separator line \citep{lau1990,priest1996}. In the absence of any current, the magnetic field in a plane orthogonal to the separator has a perpendicular X-type structure, and the reconnection was therefore originally envisaged -- projected on such a 2D plane -- as being much like 2D X-point reconnection. However, new results throw serious doubt on these simplified pictures.

Separator reconnection is perhaps the least well-understood 3D reconnection regime. One unkown factor is the extent to which the null points at either end of the separator play a significant role. Recent studies suggest that the current may preferentially peak away from the nulls \citep{parnell2010,parnell2010a}.
One thing that is clear is that the picture of cut-and-paste one-to-one rejoining of field lines at the separator line is over-simplified. If a localised current layer forms around a separator, then the reconnection within the associated diffusion region must conform to the properties described in \S\ref{fundsec}. Therefore there will be a continuous reconnection of field lines within the volume surrounding the separator. Indeed, as demonstrated by \cite{parnell2010}, the structure of the magnetic field in the vicinity of the separator may be significantly more complex than previously expected. The authors performed a simulation in which two patches of opposite polarity magnetic flux on the boundary were driven past one another \citep[see also][]{haynes2007}. While there were only two isolated, unconnected nulls in the initial configuration,  during the evolution a number of separators were formed, around which the current was found to be focussed. 
Although a separator must have a hyperbolic field near the nulls (in the plane locally perpendicular to the field), it can have either hyperbolic or elliptic structure away from the nulls.  During the simulation, the structure of the field varied both along the separator and in time as the current density fluctuated.
Clearly, reconnection at a separator with elliptic local magnetic field must be rather different to the simple early models. Indeed, \cite{parnell2010} identified the presence of counter-rotating flows around the separator on either side of localised enhancements in the parallel electric field, a signature of the non-null reconnection described in \S\ref{nonnullrec}.

\subsection{Complexity of global energy release events}
Each of the 3D reconnection regimes as discussed above is essentially a model for an isolated, local reconnection process. Two new computational studies have revealed that when a global evolution is considered there may be many of these localised reconnection events, and the evolution of the magnetic flux can be highly complex. First, we return to the scenario of two flux patches driven past one another in the presence of an overlying field, discussed in \S\ref{sepsec}. As mentioned above, reconnection is found to occur at separator lines that are formed during the simulation. Up to five separators are present during the evolution, and by measuring the temporal evolution of the reconnection rate, \cite{parnell2008} identified a `recursive reconnection' of the magnetic flux. More precisely, they found that magnetic flux was reconnected cyclically through the identified reconnection sites (separators). The result was that each unit of flux was found to be reconnected, on average, 3.6 times. This should be compared with the expectation that in an equipotential evolution each unit of flux would be reconnected twice (once to connect the two flux patches to one another, and once as they are disconnected).

Similar behaviour has been observed recently during the resistive relaxation of a magnetic flux tube containing braided magnetic flux. Beginning from an approximately force-free state, it was shown by \cite{wilmotsmith2010} that an instability caused a loss of equilibrium and the formation of a pair of current layers (in an earlier study it was shown that an {\it ideal} relaxation led towards a smooth force-free equilibrium without such current layers \citep{wilmotsmith2009a}). The subsequent evolution was investigated by \cite{pontin2011a} who found that an increasingly complex distribution of current layers formed as the relaxation progressed. This increase in complexity was eventually halted and reversed, with the final state of the relaxation being a non-linear force-free field characterised by weak, spatially distributed currents in which the flux was no longer braided. Measuring the reconnection rate associated with each of the separate current layers formed during the relaxation, it was found that each unit of flux was multiply-reconnected. The authors repeated the simulation for different magnetic Reynolds numbers, $R_M$, demonstrating that for larger $R_M$, the number of current layers increases, as does the average number of reconnections per unit of flux (the highest observed number being 4.4). The results imply that in astrophysical plasmas with very low dissipation the reconnection sites will be highly fragmented, and the global reconnection rate associated with these many linked events may be fast (even though rates for individual events may not be).

\section{3D reconnection in solar observations}\label{sunsec}
In the solar atmosphere, observed sites of energy release are often taken as signatures of a local magnetic reconnection process. Observations of the solar corona give strong indications that the continuous change of connectivity associated with 3D reconnection truly occurs there. \cite{aulanier2007} reported {\it Hinode} observations of  slippage of coronal loops, interpreted as a signature of non-null (slip-running) reconnection in a QSL. Further evidence was presented by \cite{masson2009} who described {\it TRACE (Transition Region And Coronal Explorer)} observations of propagating bright sources along a flare ribbon associated with the fan surface of a coronal null point, related to the flipping of field lines during spine-fan reconnection at the null. 

While the magnetic field in the solar corona cannot be directly measured, many studies have been performed to attempt to determine the local field structure around the reconnection site, by extrapolation from vector magnetograms. A number of these studies find the locations of energy release to be well-correlated with the locations of QSLs \citep[e.g.][]{demoulin1994,demoulin1997,mandrini2006,titov2008}. 
There are also a number of observations that suggest that null points are a common feature in flaring or eruptive locations in active regions and the quiet sun \citep[e.g.][]{mandrini2006,luoni2007,UgarteUrra2007,masson2009}. Indeed, while nulls are by no means found to be associated with all solar eruptions, a statistical study by \cite{barnes2007} showed that active regions containing nulls are more susceptible to eruption. 3D null point reconnection, via the spine-fan reconnection mode, has been proposed as a mechanism for polar jets \citep{pariat2009} as well as for CMEs via the `magnetic breakout' model \citep{antiochos1999}. 
Turning to separators, \cite{longcope2005a} have inferred the presence of separator reconnection in the corona based on observations of the emergence of a new active region in the vicinity of a pre-existing active region. One of the great difficulties in diagnosing the reconnection mechanism in 3D numerical simulations has been to determine the topology of the magnetic field in the vicinity of the reconnection site when the global field structure is complex. However, great progress has recently been made with the development of new algorithms to effectively determine the `topological skeleton' of complex 3D magnetic fields \citep{haynes2007b,haynes2010}.

\section{Reconnection in astrophysical plasmas}\label{astrosec}
The development of the reconnection theory discussed above has been motivated primarily by a desire to explain phenomena in the solar corona,  the Earth's magnetosphere, and laboratory plasmas, because these are environments that we can observe in relatively high spatial detail. Recent observations are revealing the complex structure of the magnetic field in these solar system plasmas. Reconnection is proposed to be a crucial process in many other astrophysical objects (in which it is very likely that this rich structure is also present). Therefore, the insight gained from studying processes involving reconnection within our solar system can tell us a great deal about the behaviour of more distant objects. However, one must always bear in mind that in extreme plasma environments the appropriate physics in the vicinity of the current layer may be significantly different.
Below we mention just a small number of the key roles that reconnection is proposed to play throughout the universe.

Flare events are observed in many different astrophysical objects. It is natural to suppose that these processes occur in a similar way to solar flares (though e.g.~in binary systems if the interaction of the two stars plays a crucial role this may not be the case). 
Flares -- possibly associated with CME-like eruptions -- are observed in many different kinds of stars \citep[see e.g.][for a review]{pettersen1989}, including giant flares in exotic objects like magnetars \citep{thompson2001,lyutikov2003,lyutikov2006}. It has also been suggested that reconnection may provide the energy for acceleration in jets from active galactic nuclei and gamma-ray bursts \citep{romanova1992,drenkhahn2002,giannios2010}. Furthermore, by analogy with the solar coronal heating problem, reconnection has  been put forward as a possible heating mechanism for the warm ionised medium of galaxies \citep{reynolds1999}. 

Another location in which reconnection is thought to play a key role is in accretion disks. Here we are yet to understand the process  that allows angular momentum to be lost, and thus material to accrete, in the disk \citep[for a review see][]{papaloizou1995}. One leading theory suggests that MHD turbulence in the disk leads to an effective viscosity that allows a radial diffusion of material. A key ingredient in the associated dynamo cycle is reconnection, and the value of the effective viscosity crucially depends on the efficiency of the reconnection process. \cite{coroniti1981} suggested a mechanism in which the magnetic field lies only in the plane of the disk. Strong azimuthal field is created from radial field by shearing via Keplerian motion, with some hydrodynamic turbulence in the disk driving reconnection to convert azimuthal to radial field. The origin of this turbulence, however, was not addressed in the model. \cite{tout1992} proposed a scenario by which the dynamo process can be closed self-consistently. They argued that the strongly sheared field in the plane of the disk would be susceptible to the Parker buoyancy instability, which would cause magnetic loops to rise out of the plane of the disk. With a non-zero vertical magnetic field component, the Balbus-Hawley instability may then occur, leading to reconnection of this vertical field component -- generating radial magnetic flux and thus completing the cycle. 
This reconnection of the rising vertical loops may occur in a hot corona above the disk \citep{galeev1979,uzdensky2008}.
Another proposed mechanism of angular momentum loss in such systems is via winds or jets. For example, \cite{goodson1999} have described how a magnetic interaction of the star and the disk in young stellar objects can lead, via reconnection,  to the formation of jets which carry off angular momentum both from the star and the disk \citep[see also][]{hayashi1996}. 

In ultra-relativistic pulsar winds, reconnection is invoked to explain the energy transport. Close to the pulsar, energy flux is thought to be dominated by the Poynting flux, but far away the energy must be carried mostly by  particles in order to explain observations of shock fronts associated with these winds. It is proposed that winds for oblique-rotating pulsars have a `striped' structure made up of regions of alternating-signed toroidal magnetic field, and that reconnection in the current sheets between these regions may convert magnetic to particle energy \citep[e.g.][]{coroniti1990,lyubarsky2001}. 


\section{Outlook}
Magnetic reconnection is a universal process in astrophysical plasmas. It facilitates the release of stored magnetic energy by permitting a breakdown in magnetic flux and field line conservation, and as such is a key ingredient of many energetic phenomena. In solar physics, an understanding of reconnection is crucial for gaining insight into the heating of the corona, solar flares and eruptions, and the solar dynamo. It is of great importance to understand exactly {\it how} the reconnection process facilitates these phenomena, in a plasma that we can observe with increasingly high spatial detail. While the resolution of the observations may not be sufficient to resolve the diffusion region itself, the pattern of plasma flows and locations of particle acceleration can, for example, be detected. If we can understand the underlying mechanism behind phenomena involving reconnection, then we can extrapolate this understanding to other such processes in the many disparate astrophysical systems, where such high resolution observations are unlikely to ever be possible.

While there has been great progress in understanding magnetic reconnection in recent years, there is much left to discover. Important open questions include the following.

{\it Coupling of scales:} 
\begin{itemize}
\item
How do the global dynamics of the magnetic field determine the locations of current sheet formation in complex 3D magnetic fields? Which magnetic field structures most readily host current sheets, and therefore which 3D reconnection regimes are most relevant?
\item
How are the topology of the magnetic field and external structure coupled with the physics within the diffusion region, in which two-fluid and kinetic effects may be important? In other words, how can we bridge the gap in our current understanding between 3D, MHD modelling and 2D modelling of the current layer using two-fluid and kinetic theory?
\item
How does the reconnection in turn feed back to influence the global evolution of the magnetic field? 
\end{itemize}

{\it Quantitative properties in 3D:} What are the quantitative properties of the different 3D reconnection regimes described in \S\ref{3dsec}? Thus far the vast majority of our knowledge is of a qualitative nature, and quantitative studies are required to probe, for example,  the diffusion region dimensions, the possible reconnection rates, and how these scale with different plasma parameters.

{\it Observational signatures:} What are the observational signatures of reconnection in 3D? How do they differ between the different reconnection regimes? In particular for astrophysical plasmas, it will be crucial to determine the characteristic bulk plasma flows and non-thermal particle acceleration spectra, in order to allow diagnosis via spectroscopic measurements.

{\it Local physics in the current layer:} In collisionless plasmas the structure of the electron dissipation region remains to be fully understood, as does the mechanism that determines the dimensions of this region.
In addition, more work is required to understand how the properties of reconnection are affected in extreme astrophysical environments, such as relativistic plasmas or high energy density plasmas.

{\it Energy conversion:} How is the magnetic energy transferred to other forms, and ultimately dissipated?

With the ongoing significant research effort the chances are high that the answers to many of these questions may soon be better understood, by combining the development of fundamental theory, high power simulations, and new high resolution observations.

\bibliographystyle{apalike}

\begin{thebibliography}{}

\bibitem[{Al-Hachami} and {Pontin}, 2010]{alhachami2010}
{Al-Hachami}, A.~K. and {Pontin}, D.~I. (2010).
\newblock {Magnetic reconnection at 3D null points: effect of magnetic field
  asymmetry}.
\newblock {\em Astron.~Astrophys.}, 512:A84.

\bibitem[Antiochos et~al., 1999]{antiochos1999}
Antiochos, S.~K., DeVore, C.~R., and Klimchuk, J.~A. (1999).
\newblock A model for solar coronal mass ejections.
\newblock {\em Astrophys. J.}, 510:485--493.

\bibitem[{Aulanier} et~al., 2007]{aulanier2007}
{Aulanier}, G., {Golub}, L., {DeLuca}, E.~E., {Cirtain}, J.~W., {Kano}, R.,
  {Lundquist}, L.~L., {Narukage}, N., {Sakao}, T., and {Weber}, M.~A. (2007).
\newblock {Slipping magnetic reconnection in coronal loops}.
\newblock {\em Science}, 318:1588.

\bibitem[Aulanier et~al., 2006]{aulanier2006}
Aulanier, G., Pariat, E., D{\'e}moulin, P., and Devore, C.~R. (2006).
\newblock Slip-running reconnection in quasi-separatrix layers.
\newblock {\em Solar~Phys.}, 238:347--376.

\bibitem[Barnes, 2007]{barnes2007}
Barnes, G. (2007).
\newblock On the relationship between coronal magnetic null points and solar
  eruptive events.
\newblock {\em Astrophys. J. Lett.}, 670:L53--L56.

\bibitem[{Baty} et~al., 2009a]{baty2009a}
{Baty}, H., {Forbes}, T.~G., and {Priest}, E.~R. (2009a).
\newblock {Petschek reconnection with a nonlocalized resistivity}.
\newblock {\em Phys.~Plasmas}, 16(1):012102.

\bibitem[{Baty} et~al., 2009b]{baty2009b}
{Baty}, H., {Priest}, E.~R., and {Forbes}, T.~G. (2009b).
\newblock {Petschek-like reconnection with uniform resistivity}.
\newblock {\em Phys.~Plasmas}, 16(6):060701.

\bibitem[{Bhattacharjee} et~al., 2009]{bhattacharjee2009}
{Bhattacharjee}, A., {Huang}, Y., {Yang}, H., and {Rogers}, B. (2009).
\newblock {Fast reconnection in high-Lundquist-number plasmas due to the
  plasmoid instability}.
\newblock {\em Phys.~Plasmas}, 16(11):112102.

\bibitem[Bhesso and Bhattacharjee, 2005]{bhesso2005}
Bhesso, N. and Bhattacharjee, A. (2005).
\newblock Collisionless reconnection in an electron-positron plasma.
\newblock {\em Phys.~Rev.~Lett.}, 95:245001.

\bibitem[{Birn} et~al., 2001]{birn2001}
{Birn}, J., {Drake}, J.~F., {Shay}, M.~A., {Rogers}, B.~N., {Denton}, R.~E.,
  {Hesse}, M., {Kuznetsova}, M., {Ma}, Z.~W., {Bhattacharjee}, A., {Otto}, A.,
  and {Pritchett}, P.~L. (2001).
\newblock {Geospace Environmental Modeling (GEM) magnetic reconnection
  challenge}.
\newblock {\em J.~Geophys.~Res.}, 106:3715--3720.

\bibitem[Birn and Priest, 2007]{birn2007}
Birn, J. and Priest, E.~R., editors (2007).
\newblock {\em Reconnection of Magnetic Fields : Magnetohydrodynamics and
  Collisionless Theory and Observations}.
\newblock Cambridge University Press.

\bibitem[Biskamp, 1986]{biskamp1986}
Biskamp, D. (1986).
\newblock {Magnetic reconnection via current sheets}.
\newblock {\em Phys. Fluids}, 29:1520--1531.

\bibitem[Biskamp, 2000]{biskamp2000}
Biskamp, D. (2000).
\newblock {\em Magnetic reconnection in plasmas}.
\newblock Cambridge University Press.

\bibitem[Boozer, 2005]{boozer2005}
Boozer, A.~H. (2005).
\newblock Magnetic reconnection in non-toroidal plasmas.
\newblock {\em Phys. Plasmas}, 12:070706.

\bibitem[{Browning} et~al., 2008]{browning2008}
{Browning}, P.~K., {Gerrard}, C., {Hood}, A.~W., {Kevis}, R., and {van der
  Linden}, R.~A.~M. (2008).
\newblock {Heating the corona by nanoflares: Simulations of energy release
  triggered by a kink instability}.
\newblock {\em Astron.~Astrophys.}, 485:837--848.

\bibitem[Bulanov and Sakai, 1997]{bulanov1997}
Bulanov, S.~V. and Sakai, J. (1997).
\newblock Magnetic collapse in incompressible plasma flows.
\newblock {\em J. Phys. Soc. Jpn.}, 66(11):3477--3483.

\bibitem[{Cassak} et~al., 2005]{cassak2005}
{Cassak}, P.~A., {Shay}, M.~A., and {Drake}, J.~F. (2005).
\newblock {Catastrophe model for fast magnetic reconnection onset}.
\newblock {\em Phys.~Rev.~Lett.}, 95(23):235002.

\bibitem[Coroniti, 1981]{coroniti1981}
Coroniti, F.~V. (1981).
\newblock {On the magnetic viscosity in Keplerian accretion disks}.
\newblock {\em Astrophys. J.}, 244:587--599.

\bibitem[{Coroniti}, 1990]{coroniti1990}
{Coroniti}, F.~V. (1990).
\newblock {Magnetically striped relativistic magnetohydrodynamic winds - The
  Crab Nebula revisited}.
\newblock {\em Astrophys.~J.}, 349:538--545.

\bibitem[Craig and Fabling, 1996]{craig1996}
Craig, I. J.~D. and Fabling, R.~B. (1996).
\newblock Exact solutions for steady-state, spine, and fan magnetic
  reconnection.
\newblock {\em Astrophys. J.}, 462:969--976.

\bibitem[Craig and Fabling, 1998]{craig1998}
Craig, I. J.~D. and Fabling, R.~B. (1998).
\newblock Dynamic magnetic reconnection in three space dimensions: Fan current
  solutions.
\newblock {\em Phys. Plasmas}, 5:635--644.

\bibitem[Craig et~al., 1995]{craigetal1995}
Craig, I. J.~D., Fabling, R.~B., Henton, S.~M., and Rickard, G.~J. (1995).
\newblock An exact solution for steady state magnetic reconnection in three
  dimensions.
\newblock {\em Astrophys. J. Lett.}, 455:L197--L199.

\bibitem[D\a'emoulin, 2006]{demoulin2006}
D\a'emoulin, P. (2006).
\newblock Extending the concept of separatrices to {QSLs} for magnetic
  reconnection.
\newblock {\em Adv.~Space Res.}, {37}:1269--1282.

\bibitem[D\a'emoulin et~al., 1994]{demoulin1994}
D\a'emoulin, P., H\a'enoux, J.~C., and Mandrini, C.~H. (1994).
\newblock Are null magnetic points important in solar flares?
\newblock {\em Astron. Astrophys.}, 285:1023.

\bibitem[Daughton and Karimabadi, 2007]{daughton2007}
Daughton, W. and Karimabadi, H. (2007).
\newblock Collisionless magnetic reconnection in large-scale electron-positron
  plasmas.
\newblock {\em Phys.~Plasmas}, 14:072303.

\bibitem[{Daughton} et~al., 2009]{daughton2009}
{Daughton}, W., {Roytershteyn}, V., {Albright}, B.~J., {Karimabadi}, H., {Yin},
  L., and {Bowers}, K.~J. (2009).
\newblock {Transition from collisional to kinetic regimes in large-scale
  reconnection layers}.
\newblock {\em Phys.~Rev.~Lett.}, 103(6):065004.

\bibitem[D{\'e}moulin et~al., 1997]{demoulin1997}
D{\'e}moulin, P., Bagala, L.~G., Mandrini, C.~H., H{\'e}noux, J.~C., and
  Rovira, M.~G. (1997).
\newblock Quasi-separatrix layers in solar flares. {II. Observed} magnetic
  configurations.
\newblock {\em Astron.\ Astrophys.}, 325:305--317.

\bibitem[Dorelli et~al., 2007]{dorelli2007}
Dorelli, J.~C., Bhattacharjee, A., and Raeder, J. (2007).
\newblock Separator reconnection at {Earth's} dayside magnetopause under
  generic northward interplanetary magnetic field conditions.
\newblock {\em J.~Geophys.~Res.}, 112:A02202.

\bibitem[{Drake} et~al., 2008]{drake2008}
{Drake}, J.~F., {Shay}, M.~A., and {Swisdak}, M. (2008).
\newblock {The Hall fields and fast magnetic reconnection}.
\newblock {\em Physics of Plasmas}, 15(4):042306.

\bibitem[{Drenkhahn} and {Spruit}, 2002]{drenkhahn2002}
{Drenkhahn}, G. and {Spruit}, H.~C. (2002).
\newblock {Efficient acceleration and radiation in Poynting flux powered GRB
  outflows}.
\newblock {\em Astron.~Astrophys.}, 391:1141--1153.

\bibitem[{Galeev} et~al., 1979]{galeev1979}
{Galeev}, A.~A., {Rosner}, R., and {Vaiana}, G.~S. (1979).
\newblock {Structured coronae of accretion disks}.
\newblock {\em Astrophys.~J.}, 229:318--326.

\bibitem[Galsgaard and Nordlund, 1996]{galsgaard1996}
Galsgaard, K. and Nordlund, A. (1996).
\newblock {Heating and activity of the solar corona: 1. Boundary shearing of an
  initially homogeneous magnetic field}.
\newblock {\em J. Geophys. Res.}, 101:13445--13460.

\bibitem[Galsgaard and Nordlund, 1997]{galsgaard1997}
Galsgaard, K. and Nordlund, A. (1997).
\newblock {Heating and activity of the solar corona: 3. Dynamics of a low beta
  plasma with 3D null points}.
\newblock {\em J. Geophys. Res.}, 102:231--248.

\bibitem[Galsgaard and Pontin, 2011]{galsgaard2011a}
Galsgaard, K. and Pontin, D.~I. (2011).
\newblock {Steady state reconnection at a single 3D magnetic null point}.
\newblock {\em Astron.~Astrophys.}, 529:A20.

\bibitem[Galsgaard et~al., 2003]{galsgaardpriest2003}
Galsgaard, K., Priest, E.~R., and Titov, V.~S. (2003).
\newblock {Numerical experiments on wave propagation towards a 3D null point
  due to rotational motions}.
\newblock {\em J.~Geophys.~Res.~Space}, 108:1042.

\bibitem[{Giannios}, 2010]{giannios2010}
{Giannios}, D. (2010).
\newblock {UHECRs from magnetic reconnection in relativistic jets}.
\newblock {\em Mon.\ Not.\ Roy.\ Astron.\ Soc.}, 408:L46--L50.

\bibitem[{Goodson} and {Winglee}, 1999]{goodson1999}
{Goodson}, A.~P. and {Winglee}, R.~M. (1999).
\newblock {Jets from accreting magnetic young stellar objects. II. Mechanism
  physics}.
\newblock {\em Astrophys.~J.}, 524:159--168.

\bibitem[{Hayashi} et~al., 1996]{hayashi1996}
{Hayashi}, M.~R., {Shibata}, K., and {Matsumoto}, R. (1996).
\newblock {X-ray flares and mass outflows driven by magnetic interaction
  between a protostar and its surrounding disk}.
\newblock {\em Astrophys.~J.~Lett.}, 468:L37.

\bibitem[{Haynes} and {Parnell}, 2007]{haynes2007b}
{Haynes}, A.~L. and {Parnell}, C.~E. (2007).
\newblock {A trilinear method for finding null points in a three-dimensional
  vector space}.
\newblock {\em Phys.~Plasmas}, 14(8):082107.

\bibitem[{Haynes} and {Parnell}, 2010]{haynes2010}
{Haynes}, A.~L. and {Parnell}, C.~E. (2010).
\newblock {A method for finding three-dimensional magnetic skeletons}.
\newblock {\em Phys.~Plasmas}, 17:092903.

\bibitem[{Haynes} et~al., 2007]{haynes2007}
{Haynes}, A.~L., {Parnell}, C.~E., {Galsgaard}, K., and {Priest}, E.~R. (2007).
\newblock {Magnetohydrodynamic evolution of magnetic skeletons}.
\newblock {\em Royal Society of London Proceedings Series A}, 463:1097--1115.

\bibitem[Hesse and Schindler, 1988]{hesse1988}
Hesse, M. and Schindler, K. (1988).
\newblock A theoretical foundation of general magnetic reconnection.
\newblock {\em J. Geophys. Res.}, 93:5558--5567.

\bibitem[{Hood} et~al., 2009]{hood2009}
{Hood}, A.~W., {Browning}, P.~K., and {van der Linden}, R.~A.~M. (2009).
\newblock {Coronal heating by magnetic reconnection in loops with zero net
  current}.
\newblock {\em Astron.~Astrophys.}, 506:913--925.

\bibitem[Hornig, 2001]{hornig2001}
Hornig, G. (2001).
\newblock The geometry of reconnection.
\newblock In Ricca, R.~L., editor, {\em An introduction to the geometry and
  topology of fluid flows}, pages 295--313. Kluwer, Dordrecht.

\bibitem[Hornig, 2007]{hornig2007a}
Hornig, G. (2007).
\newblock Fundamental concepts.
\newblock In Birn, J. and Priest, E.~R., editors, {\em Reconnection of magnetic
  fields}, pages 25--45. Cambridge University Press.

\bibitem[Hornig and Priest, 2003]{hornig2003}
Hornig, G. and Priest, E.~R. (2003).
\newblock Evolution of magnetic flux in an isolated reconnection process.
\newblock {\em Phys. Plasmas}, 10(7):2712--1721.

\bibitem[Hornig and Schindler, 1996]{hornig1996}
Hornig, G. and Schindler, K. (1996).
\newblock Magnetic topology and the problem of its invariant definition.
\newblock {\em Phys. Plasmas}, 3:781--791.

\bibitem[{Kliem} et~al., 2004]{kliem2004}
{Kliem}, B., {Titov}, V.~S., and {T{\"o}r{\"o}k}, T. (2004).
\newblock {Formation of current sheets and sigmoidal structure by the kink
  instability of a magnetic loop}.
\newblock {\em Astron.\ Astrophys.}, 413:L23--L26.

\bibitem[{Kliem} and {T{\"o}r{\"o}k}, 2006]{kliem2006}
{Kliem}, B. and {T{\"o}r{\"o}k}, T. (2006).
\newblock {Torus instability}.
\newblock {\em Phys.~Rev.~Lett.}, 96(25):255002.

\bibitem[Lau and Finn, 1990]{lau1990}
Lau, Y.~T. and Finn, J.~M. (1990).
\newblock Three dimensional kinematic reconnection in the presence of field
  nulls and closed field lines.
\newblock {\em Astrophys. J.}, 350:672--691.

\bibitem[Lazarian et~al., 2012]{lazarian2012}
Lazarian, A., Eyink, G. L. and Vishniac, E. T. (2012).
\newblock Relation of astrophysical turbulence and magnetic reconnection.
\newblock {\em Phys.~Plasmas}, 19:012105.

\bibitem[Linton et~al., 2001]{linton2001}
Linton, M., Dahlburg, R.~B., and Antiochos, S.~K. (2001).
\newblock Reconnection of twisted flux tubes as a function of contact angle.
\newblock {\em Astrophys. J.}, 553:905--921.

\bibitem[Longcope, 1996]{longcope1996}
Longcope, D.~W. (1996).
\newblock Topology and current ribbons: A model for current, reconnection and
  flaring in a complex, evolving corona.
\newblock {\em Solar Phys.}, 169:91--121.

\bibitem[Longcope and Cowley, 1996]{longcopecowley1996}
Longcope, D.~W. and Cowley, S.~C. (1996).
\newblock Current sheet formation along three-dimensional magnetic separators.
\newblock {\em Phys. Plasmas}, 3:2885--2897.

\bibitem[Longcope et~al., 2005]{longcope2005a}
Longcope, D.~W., McKenzie, D., Cirtain, J., and Scott, J. (2005).
\newblock Observations of separator reconnection to an emerging active region.
\newblock {\em Astrophys. J.}, 630:569.

\bibitem[{Longcope} and {Parnell}, 2009]{longcope2009}
{Longcope}, D.~W. and {Parnell}, C.~E. (2009).
\newblock {The number of magnetic null points in the quiet sun corona}.
\newblock {\em Solar Phys.}, 254:51--75.

\bibitem[{Longcope} and {Strauss}, 1994]{longcope1994}
{Longcope}, D.~W. and {Strauss}, H.~R. (1994).
\newblock {The form of ideal current layers in line-tied magnetic fields}.
\newblock {\em Astrophys.~J.}, 437:851--859.

\bibitem[{Loureiro} et~al., 2009]{loureiro2009}
{Loureiro}, N.~F., {Uzdensky}, D.~A., {Schekochihin}, A.~A., {Cowley}, S.~C.,
  and {Yousef}, T.~A. (2009).
\newblock {Turbulent magnetic reconnection in two dimensions}.
\newblock {\em Mon.\ Not.\ Roy.\ Astron.\ Soc.}, 399:L146--L150.

\bibitem[Luoni et~al., 2007]{luoni2007}
Luoni, M.~L., Mandrini, H.~H., Cristiani, G.~D., and D{\' e}moulin, P. (2007).
\newblock The magnetic field topology associated with two {M} flares.
\newblock {\em Adv. Space Res.}, 39:1382--1388.

\bibitem[{Lyubarsky} and {Kirk}, 2001]{lyubarsky2001}
{Lyubarsky}, Y. and {Kirk}, J.~G. (2001).
\newblock {Reconnection in a striped pulsar wind}.
\newblock {\em Astrophys.~J.}, 547:437--448.

\bibitem[{Lyutikov}, 2003]{lyutikov2003}
{Lyutikov}, M. (2003).
\newblock {Explosive reconnection in magnetars}.
\newblock {\em Mon.~Not.~Roy.~Astron.~Soc.}, 346:540--554.

\bibitem[{Lyutikov}, 2006]{lyutikov2006}
{Lyutikov}, M. (2006).
\newblock {Magnetar giant flares and afterglows as relativistic magnetized
  explosions}.
\newblock {\em Mon.~Not.~Roy.~Astron.~Soc.}, 367:1594--1602.

\bibitem[Mandrini et~al., 2006]{mandrini2006}
Mandrini, C.~H., D{\' e}moulin, P., Schmieder, B., Deluca, E.~E., Pariat, E.,
  and Uddin, W. (2006).
\newblock Companion event and precursor of the {X17} flare on 28 {October}
  2003.
\newblock {\em Solar Phys.}, 238:293--312.

\bibitem[Masson et~al., 2009]{masson2009}
Masson, S., Pariat, E., Aulanier, G., and Schrijver, C.~J. (2009).
\newblock The nature of flare ribbons in coronal null-point topology.
\newblock {\em Astrophys.\ J.}, 700:559--578.

\bibitem[McLaughlin et~al., 2011]{mclaughlin2011}
McLaughlin, J.~A., Hood, A.~W., and de~Moortel, I. (2011).
\newblock Review article: {MHD} wave propagation near coronal null points of
  magnetic fields.
\newblock {\em Space~Sci.~Rev.}, 158:205--236.

\bibitem[{Papaloizou} and {Lin}, 1995]{papaloizou1995}
{Papaloizou}, J.~C.~B. and {Lin}, D.~N.~C. (1995).
\newblock {Theory of accretion disks I: Angular momentum transport processes}.
\newblock {\em Ann.\ Rev.\ Astron.\ Astrophys.}, 33:505--540.

\bibitem[{Pariat} et~al., 2009]{pariat2009}
{Pariat}, E., {Antiochos}, S.~K., and {DeVore}, C.~R. (2009).
\newblock {A model for solar polar jets}.
\newblock {\em Astrophys. J.}, 691:61--74.

\bibitem[{Parker}, 1957]{parker1957}
{Parker}, E.~N. (1957).
\newblock {Sweet's mechanism for merging magnetic fields in conducting fluids}.
\newblock {\em J.~Geophys.~Res.}, 62:509--520.

\bibitem[Parnell et~al., 2008]{parnell2008}
Parnell, C.~E., Haynes, A.~L., and Galsgaard, K. (2008).
\newblock Recursive reconnection and magnetic skeletons.
\newblock {\em Astrophys.~J.}, 675:1656--1667.

\bibitem[Parnell et~al., 2010a]{parnell2010}
Parnell, C.~E., Haynes, A.~L., and Galsgaard, K. (2010a).
\newblock Structure of magnetic separators and separator reconnection.
\newblock {\em J.~Geophys.~Res.}, 115:A02102.

\bibitem[Parnell et~al., 2010b]{parnell2010a}
Parnell, C.~E., MacLean, R.~C., and Haynes, A.~L. (2010b).
\newblock The detection of numerous magnetic separators in a three-dimensional
  magnetohydrodynamic model of solar emerging flux.
\newblock {\em Astrophys.~J.~Lett.}, 725:214--218.

\bibitem[Parnell et~al., 1996]{parnell1996}
Parnell, C.~E., Smith, J.~M., Neukirch, T., and Priest, E.~R. (1996).
\newblock The structure of three-dimensional magnetic neutral points.
\newblock {\em Phys.~Plasmas}, 3(3):759--770.

\bibitem[Petschek, 1964]{petschek1964}
Petschek, H.~E. (1964).
\newblock {\em Magnetic field annihilation}, pages 425--439.
\newblock NASA SP-50, Washington, D.C.

\bibitem[{Pettersen}, 1989]{pettersen1989}
{Pettersen}, B.~R. (1989).
\newblock {A review of stellar flares and their characteristics}.
\newblock {\em Solar Phys.}, 121:299--312.

\bibitem[Pontin, 2011]{pontin2011b}
Pontin, D.~I. (2011).
\newblock Three-dimensional magnetic reconnection regimes: A review.
\newblock {\em Adv.~Space Res.}, 47:1508--1522.

\bibitem[Pontin et~al., 2007a]{pontinbhat2007a}
Pontin, D.~I., Bhattacharjee, A., and Galsgaard, K. (2007a).
\newblock Current sheet formation and non-ideal behaviour at three-dimensional
  magnetic null points.
\newblock {\em Phys.~Plasmas}, 14:052106.

\bibitem[Pontin et~al., 2007b]{pontinbhat2007b}
Pontin, D.~I., Bhattacharjee, A., and Galsgaard, K. (2007b).
\newblock Current sheets at three-dimensional magnetic null points: Effect of
  compressibility.
\newblock {\em Phys.~Plasmas}, 14:052109.

\bibitem[Pontin and Craig, 2005]{pontincraig2005}
Pontin, D.~I. and Craig, I.~J.~D. (2005).
\newblock Current singularities at finitely compressible three-dimensional
  magnetic null points.
\newblock {\em Phys.~Plasmas}, 12:072112.

\bibitem[Pontin and Galsgaard, 2007]{pontingalsgaard2007}
Pontin, D.~I. and Galsgaard, K. (2007).
\newblock {Current amplification and magnetic reconnection at a 3D null point.
  Physical characteristics}.
\newblock {\em J.~Geophys.~Res.}, 112:A03103.

\bibitem[Pontin et~al., 2005a]{pontingalsgaard2005}
Pontin, D.~I., Galsgaard, K., Hornig, G., and Priest, E.~R. (2005a).
\newblock {A fully magnetohydrodynamic simulation of 3D non-null reconnection}.
\newblock {\em Phys.~Plasmas}, 12:052307.

\bibitem[Pontin et~al., 2004]{pontin2004}
Pontin, D.~I., Hornig, G., and Priest, E.~R. (2004).
\newblock Kinematic reconnection at a magnetic null point: Spine-aligned
  current.
\newblock {\em Geophys. Astrophys. Fluid Dynamics}, 98:407--428.

\bibitem[Pontin et~al., 2005b]{pontinhornig2005}
Pontin, D.~I., Hornig, G., and Priest, E.~R. (2005b).
\newblock Kinematic reconnection at a magnetic null point: Fan-aligned current.
\newblock {\em Geophys. Astrophys. Fluid Dynamics}, 99:77--93.

\bibitem[Pontin et~al., 2011]{pontin2011a}
Pontin, D.~I., Wilmot-Smith, A.~L., Hornig, G., and Galsgaard, K. (2011).
\newblock {Dynamics of braided coronal loops. II. Cascade to multiple
  small-scale reconnection events}.
\newblock {\em Astron.~Astrophys.}, 525:A57.

\bibitem[Priest and D\a'emoulin, 1995]{priest1995}
Priest, E.~R. and D\a'emoulin, P. (1995).
\newblock {Three-dimensional magnetic reconnection without null points. 1.
  Basic theory of magnetic flipping}.
\newblock {\em J. Geophys. Res.}, 100:23443--23463.

\bibitem[Priest and Forbes, 1992]{priest1992}
Priest, E.~R. and Forbes, T.~G. (1992).
\newblock Magnetic flipping - reconnection in three dimensions without null
  points.
\newblock {\em J. Geophys. Res.}, 97:1521--1531.

\bibitem[Priest and Forbes, 2000]{priest2000}
Priest, E.~R. and Forbes, T.~G. (2000).
\newblock {\em Magnetic reconnection: MHD theory and applications}.
\newblock Cambridge University Press, Cambridge.

\bibitem[Priest et~al., 2003]{priesthornig2003}
Priest, E.~R., Hornig, G., and Pontin, D.~I. (2003).
\newblock On the nature of three-dimensional magnetic reconnection.
\newblock {\em J. Geophys. Res.}, 108(A7):1285.

\bibitem[Priest and Pontin, 2009]{priest2009}
Priest, E.~R. and Pontin, D.~I. (2009).
\newblock Three-dimensional null point reconnection regimes.
\newblock {\em Phys.~Plasmas}, 16:122101.

\bibitem[Priest and Titov, 1996]{priest1996}
Priest, E.~R. and Titov, V.~S. (1996).
\newblock Magnetic reconnection at three-dimensional null points.
\newblock {\em Phil. Trans. R. Soc. A}, 354:2951--2992.

\bibitem[Rappazzo et~al., 2008]{rappazzo2008}
Rappazzo, A.~F., Velli, M., Einaudi, G., and Dahlburg, R.~B. (2008).
\newblock {Nonlinear dynamics of the Parker scenario for coronal heating}.
\newblock {\em Astrophys. J.}, 677:1348--1366.

\bibitem[{R{\'e}gnier} et~al., 2008]{regnier2008}
{R{\'e}gnier}, S., {Parnell}, C.~E., and {Haynes}, A.~L. (2008).
\newblock {A new view of quiet-Sun topology from Hinode/SOT}.
\newblock {\em Astron. Astrophys.}, 484:L47--L50.

\bibitem[{Reynolds} et~al., 1999]{reynolds1999}
{Reynolds}, R.~J., {Haffner}, L.~M., and {Tufte}, S.~L. (1999).
\newblock {Evidence for an additional heat source in the warm ionized medium of
  galaxies}.
\newblock {\em Astrophys.~J.~Lett.}, 525:L21--L24.

\bibitem[Rickard and Titov, 1996]{rickard1996}
Rickard, G.~J. and Titov, V.~S. (1996).
\newblock Current accumulation at a three-dimensional magnetic null.
\newblock {\em Astrophys. J.}, 472:840--852.

\bibitem[Romanova and Lovelace, 1992]{romanova1992}
Romanova, M.~M. and Lovelace, R. V.~E. (1992).
\newblock Magnetic field reconnection and particle acceleration in
  extragalactic jets.
\newblock {\em Astron. Astrophys.}, 262:26--36.

\bibitem[Schindler et~al., 1988]{schindler1988}
Schindler, K., Hesse, M., and Birn, J. (1988).
\newblock General magnetic reconnection, parallel electric fields, and
  helicity.
\newblock {\em J.~Geophys.~Res.}, 93(A6):5547--5557.

\bibitem[Sweet, 1958]{sweet1958}
Sweet, P.~A. (1958).
\newblock The neutral point theory of solar flares.
\newblock In {\em Electromagnetic phenomena in cosmical plasmas}, pages
  123--134. Cambridge University Press, Cambridge.

\bibitem[Syrovatskii, 1971]{syrovatskii1971}
Syrovatskii, S.~I. (1971).
\newblock Formation of current sheets in a plasma with frozen-in strong
  magnetic field.
\newblock {\em Sov. Phys. JETP}, 33:933--940.

\bibitem[{Thompson} and {Duncan}, 2001]{thompson2001}
{Thompson}, C. and {Duncan}, R.~C. (2001).
\newblock {The giant flare of 1998 August 27 from SGR 1900+14. II. Radiative
  mechanism and physical constraints on the source}.
\newblock {\em Astrophys.~J.}, 561:980--1005.

\bibitem[Titov, 2007]{titov2007}
Titov, V.~S. (2007).
\newblock Generalized squashing factors for covariant description of magnetic
  connectivity in the solar corona.
\newblock {\em Astrophys.\ J.}, 660:863--873.

\bibitem[Titov et~al., 2002]{titov2002}
Titov, V.~S., Hornig, G., and D\a'emoulin, P. (2002).
\newblock The theory of magnetic connectivity in the corona.
\newblock {\em J. Geophys. Res.}, 107:1164.

\bibitem[{Titov} et~al., 2008]{titov2008}
{Titov}, V.~S., {Mikic}, Z., {Linker}, J.~A., and {Lionello}, R. (2008).
\newblock {1997 May 12 coronal mass ejection event. I. A simplified model of
  the preeruptive magnetic structure}.
\newblock {\em Astrophys.\ J.}, 675:1614--1628.

\bibitem[Tout and Pringle, 1992]{tout1992}
Tout, C. and Pringle, J.~E. (1992).
\newblock Accretion disk viscosity - a simple model for a magnetic dynamo.
\newblock {\em Mon. Not. Roy. Astron. Soc.}, 259:604--612.

\bibitem[Ugarte-Urra et~al., 2007]{UgarteUrra2007}
Ugarte-Urra, I., Warren, H.~P., and Winebarger, A.~R. (2007).
\newblock The magnetic topology of coronal mass ejection sources.
\newblock {\em Astrophys. J.}, 662:1293--1301.

\bibitem[{Uzdensky} and {Goodman}, 2008]{uzdensky2008}
{Uzdensky}, D.~A. and {Goodman}, J. (2008).
\newblock Statistical description of a magnetized corona above a turbulent
  accretion disk.
\newblock {\em Astrophys.~J.}, 682:608--629.

\bibitem[{Uzdensky} et~al., 2010]{uzdensky2010}
{Uzdensky}, D.~A., {Loureiro}, N.~F., and {Schekochihin}, A.~A. (2010).
\newblock Fast magnetic reconnection in the plasmoid-dominated regime.
\newblock {\em Phys.~Rev.~Lett.}, 105(23):235002.

\bibitem[Wilmot-Smith et~al., 2009]{wilmotsmith2009a}
Wilmot-Smith, A.~L., Hornig, G., and Pontin, D.~I. (2009).
\newblock Magnetic braiding and parallel electric fields.
\newblock {\em Astrophys. J.}, 696:1339--1347.

\bibitem[Wilmot-Smith et~al., 2010]{wilmotsmith2010}
Wilmot-Smith, A.~L., Pontin, D.~I., and Hornig, G. (2010).
\newblock {Dynamics of braided coronal loops - I. Onset of magnetic
  reconnection}.
\newblock {\em Astron. Astrophys.}, 516:A5.

\bibitem[{Xiao} et~al., 2007]{xiao2007}
{Xiao}, C.~J., {Wang}, X.~G., {Pu}, Z.~Y., {Ma}, Z.~W., {Zhao}, H., {Zhou},
  G.~P., {Wang}, J.~X., {Kivelson}, M.~G., {Fu}, S.~Y., {Liu}, Z.~X., {Zong},
  Q.~G., {Dunlop}, M.~W., {Glassmeier}, K., {Lucek}, E., {Reme}, H.,
  {Dandouras}, I., and {Escoubet}, C.~P. (2007).
\newblock {Satellite observations of separator-line geometry of
  three-dimensional magnetic reconnection}.
\newblock {\em Nature Physics}, 3:609--613.

\bibitem[{Yamada} et~al., 2010]{yamada2010}
{Yamada}, M., {Kulsrud}, R., and {Ji}, H. (2010).
\newblock {Magnetic reconnection}.
\newblock {\em Rev.~Modern~Phys.}, 82:603--664.

\bibitem[{Zweibel} and {Yamada}, 2009]{zweibel2009}
{Zweibel}, E.~G. and {Yamada}, M. (2009).
\newblock Magnetic reconnection in astrophysical and laboratory plasmas.
\newblock {\em Ann.\ Rev.\ Astron.\ Astrophys.}, 47:291--332.

\end{thebibliography}

\end{document}